\documentclass[12pt]{article}
\pdfoutput=1
\usepackage{amssymb,amsmath,amsthm,trig,amsfonts,graphicx}
\usepackage{relsize,ccaption,url,latexsym,epsfig,a4wide}
\usepackage{ccaption,url,latexsym,epsfig,a4wide}

\usepackage{appendix}

\numberwithin{equation}{section}

\captionnamefont{\bfseries}

\title{Giant magnons and non-maximal giant gravitons}

\author{\\ \Large{A. Ciavarella\footnote{a.m.ciavarella@durham.ac.uk}}\\ \\ \\
 \normalsize{Centre for Particle Theory}\\ \normalsize{Department of Mathematical Sciences} \\ \normalsize{University of Durham} \\ \normalsize{Durham, DH1 3LE, U.K.  }}

\newcommand{\be}{\begin{equation}}
\newcommand{\ee}{\end{equation}}
\newcommand{\bea}{\begin{eqnarray}}
\newcommand{\eea}{\end{eqnarray}}
\newcommand{\Ra}{\Rightarrow}
\newcommand{\nn}{\nonumber}

\newcommand{\nl}{\newline}

\begin{document}

\maketitle

\abstract{We produce the open strings on $\mathbb{R}\times S^{2}$ that correspond to the solutions of integrable boundary sine-Gordon theory by making use of the $N$-magnon solutions provided in \cite{KPV} together with explicit moduli.  Relating the two boundary parameters in a special way we describe the scattering of giant magnons with non-maximal $Y=0$ giant gravitons and calculate the leading contribution to the associated magnon scattering phase.}

\section{Introduction}

In recent work on the $AdS$/CFT correspondence\cite{Maldacena:1997re} relating type IIB strings on $AdS_{5}\times S^{5}$ to $\mathcal{N}=4$ $SU(N)$ super Yang-Mills on $\mathbb{R}^{1,3}$ particular use has been made of spinning string solutions\cite{Gubser:2002tv} which provide a niche of the duality in which exact expressions may be obtained for quantities such as the anomalous dimension of gauge theory operators at strong 't Hooft coupling $\lambda$.

The string theory description of states in the correspondence reduces to supergravity at large $\lambda$, becoming classical at large $N$, where the gauge theory is then planar.  Another feature of the large $N$ limit is the possibility of composing single trace gauge invariant operators of large $R$ charge that describe single particle string states, individual fundamental closed strings, of large angular momentum on $S^{5}$\cite{Berenstein:2002jq}, the appearance of an effective 1+1 dimensional field theory then making explicit the presence a string world sheet within non-abelian gauge theory together with a convenient new set of quantities in which to expand.

The discovery of integrability\cite{Minahan:2002ve,Arutyunov:2004vx} in the large $N$ limit, expected to hold for arbitrary $\lambda$, promised an approach to solving the theory in this limit in terms of the scattering of single excitations about the `BMN vacuum' provided by large $R$ charge scalar operators.  The anomalous dimensions of these excitations, or the dispersion relation for the corresponding strings, could be fixed on the grounds of the (enhanced) supersymmetry of these states alone\cite{Beisert:2005tm} and the scattering matrix constructed up to an overall phase factor.  Specifically it was shown that one can reduce the problem of finding the anomalous dimensions $\Delta-J$ of nearly BMN Yang-Mills operators to the problem of diagonalising an integrable $SO(6)$ spin-chain of length $J$ whose `magnon' excitations with spin-chain momentum $p$ obey
\be
\Delta-J=\sqrt{1+\frac{\lambda}{\pi^{2}}{\rm sin}^{2}\frac{p}{2}}.\label{DeltaMinusJ}
\ee
Both $\Delta$ and $J$ are strictly divergent while the difference is finite.  

In the string picture $\Delta$ becomes the string energy $E$ while the $R$ charge $J$ is the angular momentum of the string on $S^{5}$.  The string corresponding to the BMN operator with $\Delta-J=0$ is point-like and moves around a great circle on $S^{5}$ at the speed of light.  In \cite{Hofman:2006xt} the strings dual to the magnons were identified and these giant excitations around the BMN string, the giant magnons, possessing net world sheet momentum $p$ were shown to obey
\be
E-J=\frac{\sqrt{\lambda}}{\pi}\Big|{\rm sin}\frac{p}{2}\Big|,\quad N\to\infty,~\lambda~{\rm large,~fixed}, 
\ee
which is seen to be the large $\lambda$ limit of \eqref{DeltaMinusJ}.  The shape of a giant magnon is a semi-circular arc with its two ends moving at the speed of light around a great circle of $S^{5}$ so that the string interpolates between two BMN vacua.  A giant magnon is not a closed string on its own but many of these string segments may be joined end to end to close the string and the gauge theory equivalent is the demand that we trace over the long product of operators to achieve gauge invariance.

A great deal of generalisations followed including giant magnons with multiple angular momenta\cite{Chen:2006gea}, finite J\cite{Arutyunov:2006gs} and multi-magnon solutions\cite{Spradlin,KPV}.

Giant gravitons\cite{McGreevy:2000cw} are stable but finite energy D-branes on the $AdS_{5}\times S^{5}$ background that may act as end points for the open strings that emerge as their excitations\cite{Balasubramanian:2002sa}.  In gauge theory giant gravitons are given by determinants and sub-determinants of scalar operators which provide for a maximum $R$ charge that on the string side is naturally due to the upper limit on the size of the graviton on the compact part of its target space.  The giants of maximum and less than maximum size are referred to as maximal and non-maximal respectively. The giant gravitons, which have angular momenta of order $N$, wrap an $S^{3}\subset S^{5}$ and remain spherical and heavy as long the strings attached to them have angular momenta of order $\sqrt{N}$ for which the planar description of the string in gauge theory is respected.

Giant magnons were then considered upon genuinely open strings attached to giant gravitons\cite{HofmanII} and calculations of the magnon boundary scattering matrices at weak coupling was possible along the lines of \cite{Beisert:2005tm}.  Again, the (boundary) scattering matrices $\mathcal{R}$ could be determined only up to a phase, $\mathcal{R}_{0}=e^{i\delta}$,
\be
\mathcal{R}=\mathcal{R}_{0}\hat{\mathcal{R}},
\ee
which could however be calculated to leading order at large $\lambda$, albeit for boundary conditions corresponding to the presence of maximal giant gravitons only, by resort to the dual description of the giant magnons by sine-Gordon theory.

Sine-Gordon theory appears upon taking the Pohlmeyer reduction\cite{Pohlmeyer:1975nb,Hollowood:2009tw} of the string sigma model on $\mathbb{R}\times S^{2}\subset AdS_{5}\times S^{5}$ at large $\lambda$ and consists a reformulation of the theory in terms of quantities invariant under the isometries of the target space. The result is a one-to-one map at the classical level\footnote{The Poisson bracket structure of the two theories are different and so quantisation proceeds differently in each case, although for efforts toward extending the application of Pohlmeyer reduction to the quantum string theory see \cite{Hoare:2009rq}.} between the solutions of sine-Gordon theory and string solutions.  The kinks and anti-kinks of the sine-Gordon field $\varphi(x,t)$ map in a non-trivial manner to the giant magnon solutions $X^{i}(x,t)$ via
\bea
{\rm cos}\varphi=\dot{X}^{i}\dot{X}^{i}-X'^{i}X'^{i},\qquad X^{i}\in \mathbb{R}^{3},\\
{\rm or}\qquad {\rm sin}^{2}\frac{\varphi}{2}=X'^{i}X'^{i},\qquad {\rm cos}^{2}\frac{\varphi}{2}=\dot{X}^{i}\dot{X}^{i}.\label{SGDef}
\eea
from which it can be seen that one of the Virasoro constraints on the string,
\be
\dot{X}^{i}\dot{X}^{i}+X'^{i}X'^{i}=1,
\ee
is immediately satisfied.  The space time $\mathbb{R}^{1,1}$ of sine-Gordon theory is exactly the world sheet of the string and so time differences in each picture are identical.  In particular the time delays for scattering processes are identical and as explained in section 4 this leads us to semi-classical results for the scattering phase delay $\delta$.

For open strings we require boundary sine-Gordon theory which may be formulated to preserve the integrability of the bulk theory\cite{SSW}.  Localisations of energy at the integrable boundary in sine-Gordon theory appear in the string picture as boundary giant magnons\cite{HofmanII,Bak:2008xq,Ciavarella:2010je} which connect the brane to the `BMN vacuum' at the equator.  Any giant magnon / brane scattering described by an integrable boundary field theory on the world sheet will return giant magnons in tact after scattering and a boundary scattering phase delay $\delta_{B}$ may be considered.

The boundary conditions for which the authors of \cite{HofmanII} obtained their results fell into the class of boundary conditions for which the field theory and the string theory are known to be integrable\cite{Ghoshal,Mann:2006rh,MacKay:2010ey}.  The analysis was restricted to giant magnons and \emph{maximal} giant gravitons however.  It will be a result of this paper that a vestige of integrability remains on the string side at strong coupling for at least a certain ($Y=0$) non-maximal giant graviton.
\nl

This paper is organised as follows.  In section 2 we review the relevant aspects of boundary sine-Gordon theory and the method of images, highlighting the asymptotic behaviour of the boundary scattering solution.  In section 3 we describe the 3-soliton giant magnon solution with moduli and the appearance of the sine-Gordon boundary parameters upon application of the method images and reduction of the solution from $\mathbb{R}\times S^{3}$ to $\mathbb{R}\times S^{2}$.  Section 4 presents the string solutions for scattering giant magnons and giant gravitons including three maximal cases and an interesting non-maximal case, including calculations of the leading magnon boundary scattering phase.  Section 5 is the discussion.

\section{Boundary sine-Gordon theory}

In order to discuss the one-to-one map between string solutions on $\mathbb{R}\times S^{2}$ and those of boundary sine-Gordon theory we review its construction and solutions, paying heed to the asymptotic behaviour that will later be used to match the two sets of solutions.

Sine-Gordon theory for a single scalar field $\varphi(x,t)\in\mathbb{R}$ may be defined on the half-line\cite{SSW} while preserving its integrability, i.e. conserving an infinite number of constants of motion and $N$-soliton solutions.  The Lagrangian is modified by the inclusion of a boundary potential\cite{Ghoshal}, conventionally evaluated at $x=0$, specifically
\be
\mathcal{L}=\frac{1}{2}\left\{\dot{\varphi}^{2}-\varphi'^{2}+g{\rm cos}(\beta\varphi)\right\}+M{\rm cos}\left(\frac{\beta(\varphi-\varphi_{0})}{2}\right)\delta(x),
\ee
and then taking either $x\in(-\infty,0]$ or $x\in[0,\infty)$. Rescaling the field and coupling constant, and varying the action, then without loss of generality we may obtain the equation of motion
\be
\ddot{\varphi}-\varphi''=-{\rm sin}(\varphi)
\ee
plus the boundary condition
\be
\varphi'|_{x=0}=M{\rm sin}\left(\frac{\varphi-\varphi_{0}}{2}\right)\Big|_{x=0}.\label{IBC}
\ee

Solutions to the boundary theory may then be constructed by the method of images, whereby a three soliton solution to the bulk theory suffices to describe the in-going soliton, the out-going (reflected) soliton and a non-trivial boundary.  We will follow the notation of \cite{SSW}.

The $N$-soliton solution to the bulk theory is expressed through the $\tau$-function as
\be
\varphi(x,t)=4{\rm arctan}\left(\frac{\mathcal{I}(\tau)}{\mathcal{R}(\tau)}\right),\qquad \tau\equiv\tau(x,t)\in\mathbb{C}
\ee
where the $\tau$-function itself is
\bea
\tau=\sum_{\mu_{j}=0,1}e^{-\frac{\pi}{2}\left(\sum^{N}_{j=1}\epsilon_{j}\mu_{j}\right)}{\rm exp}\left\{-\sum_{j=1}^{N}\mu_{j}\left[{\rm cosh}(\theta_{j})x+{\rm sinh}(\theta_{j})t+2a_{j}\right]\right.\\
\left. +2\sum_{1\leq i<j\leq N}\mu_{i}\mu_{j}{\rm ln}\left({\rm tanh}\left(\frac{\theta_{i}-\theta_{j}}{2}\right)\right)\right\}.\hspace{1cm}
\eea
This describes a mix of $N$ kinks and / or anti-kinks, which are topologically stable solutions interpolating between the vacua of the theory at $\varphi=2m\pi,~m\in\mathbb{Z}$. 

The rapidity of the $j$th soliton at early and late times is $\theta_{j}$, the $a_{j}$ are initial positions and the $\epsilon_{j}=\pm 1$ determine whether the $j$th soliton is a kink or an anti-kink.

To perform the method of images we must take two solitons to have equal and opposite velocities and one to be stationary, therefore we take
\be
\theta_{1}\equiv\theta,\quad \theta_{3}=-\theta,\quad \theta_{2}=0
\ee
and with the definitions
\be
\epsilon\equiv\epsilon_{1}\epsilon_{3},\quad\epsilon_{0}\equiv\epsilon_{2},\quad a_{+}=a_{1}+a_{3},\quad a_{-}=a_{1}-a_{3},\quad b\equiv a_{2}
\ee
\be
v={\rm tanh}(\theta),\quad \gamma={\rm cosh}(\theta)
\ee
we construct the $\tau$ function appropriate to the method of images as
\bea
\tau=1-\epsilon v^{2}e^{-2\gamma x-a_{+}}-\epsilon_{0}\kappa^{2}e^{-(\gamma+1)x-b}F(t)\hspace{3cm}\nn\\
\hspace{3cm}+i\left\{e^{-\gamma x}F(t)+\epsilon_{0}e^{-x-b}-\epsilon\epsilon_{0}v^{2}\kappa^{4}e^{-(2\gamma+1)x-a_{+}-b}\right\}\label{MOISol}
\eea
where we have defined
\be
\kappa\equiv{\rm tanh}\left(\frac{\theta}{2}\right).
\ee
The time dependence is confined to the factor $F(t)$,
\bea
F(t)&=&\epsilon_{1}e^{-v\gamma t-a_{1}}+\epsilon_{3}e^{v\gamma t-a_{3}}\\
&=&\epsilon_{1}e^{-\frac{1}{2}a_{+}}\left(e^{-v\gamma t-\frac{1}{2}a_{-}}+\epsilon e^{v\gamma t+\frac{1}{2}a_{-}}\right)
\eea
where the second form shows us that of the 3 parametric degrees of freedom entering via the initial positions $a_{j}$ we may remove one by an over all translation in time of $t\to t-\frac{1}{2v\gamma}a_{-}$.

The integrable boundary condition is determined by the two real parameters $M$ and $\varphi_{0}$, which may be translated into the two parameters inherited as initial positions, $a_{+}$ and $b$. The three discrete parameters $\epsilon_{1}$, $\epsilon$ and $\epsilon_{0}$ are fixed by the choice boundary conditions at both $x=0$ and $x\to \pm\infty$. 

The asymptotic behaviour of the solution on $x\in [0,\infty)$ consists an in-going left-mover at $t\to-\infty$ and an out-going right-mover at $t\to+\infty$ while the elastic boundary returns to its initial state. To follow the left-mover we may take $x\to+\infty$, $t\to-\infty$ with $\zeta_{+}=\gamma(x+vt)$ fixed, leading to
\be
{\rm tan}\left(\frac{\varphi}{4}\right)\to\epsilon_{2}e^{-\zeta_{+}-a_{3}},\quad{\rm or}\quad {\rm sin}\left(\frac{\varphi}{2}\right)\to\frac{\epsilon_{2}}{{\rm cosh}(\zeta_{+}+a_{3})}
\ee
and to follow the right-mover take $x\to+\infty$, $t\to+\infty$ with $\zeta_{-}=\gamma(x-vt)$ fixed leading to
\be
{\rm tan}\left(\frac{\varphi}{4}\right)\to\epsilon_{1}e^{-\zeta_{-}-a_{1}},\quad{\rm or}\quad {\rm sin}\left(\frac{\varphi}{2}\right)\to\frac{\epsilon_{1}}{{\rm cosh}(\zeta_{-}+a_{1})}.
\ee
We recognise these as single kink or anti-kink solutions proceeding left and right respectively, with initial positions $a_{3}$ and $a_{1}$.  Meanwhile the asymptotic behaviour of the central soliton is found at $t\to\pm\infty$ by taking $x$ fixed, giving
\be
{\rm tan}\left(\frac{\varphi}{4}\right)\to-\epsilon_{0}e^{x+b-2{\rm ln}(\kappa)},\quad{\rm or}\quad {\rm sin}\left(\frac{\varphi}{2}\right)\to\frac{-\epsilon_{0}}{{\rm cosh}(x+b-2{\rm ln}(\kappa))}.
\ee
Note that the position of the central soliton is given not just by the parameter $b$ but depends upon the speed of the incoming soliton as $x_{0}=2{\rm ln}(\kappa)-b$.  The parameters $b$ and $\varphi_{0}$ are related by
\be
b={\rm ln}\left[-\epsilon_{0}\kappa^{2}{\rm tan}\left(\frac{\varphi_{0}}{4}\right)\right].\label{bphi0}
\ee
We see that $\epsilon_{0}=\pm 1$ must be picked such that $-\epsilon_{0}{\rm tan}\left(\frac{\varphi_{0}}{4}\right)$ is positive and so is fixed by the boundary condition at $x=0$.

\begin{figure}[!h]
\begin{center}
\includegraphics[width=0.4\textwidth]{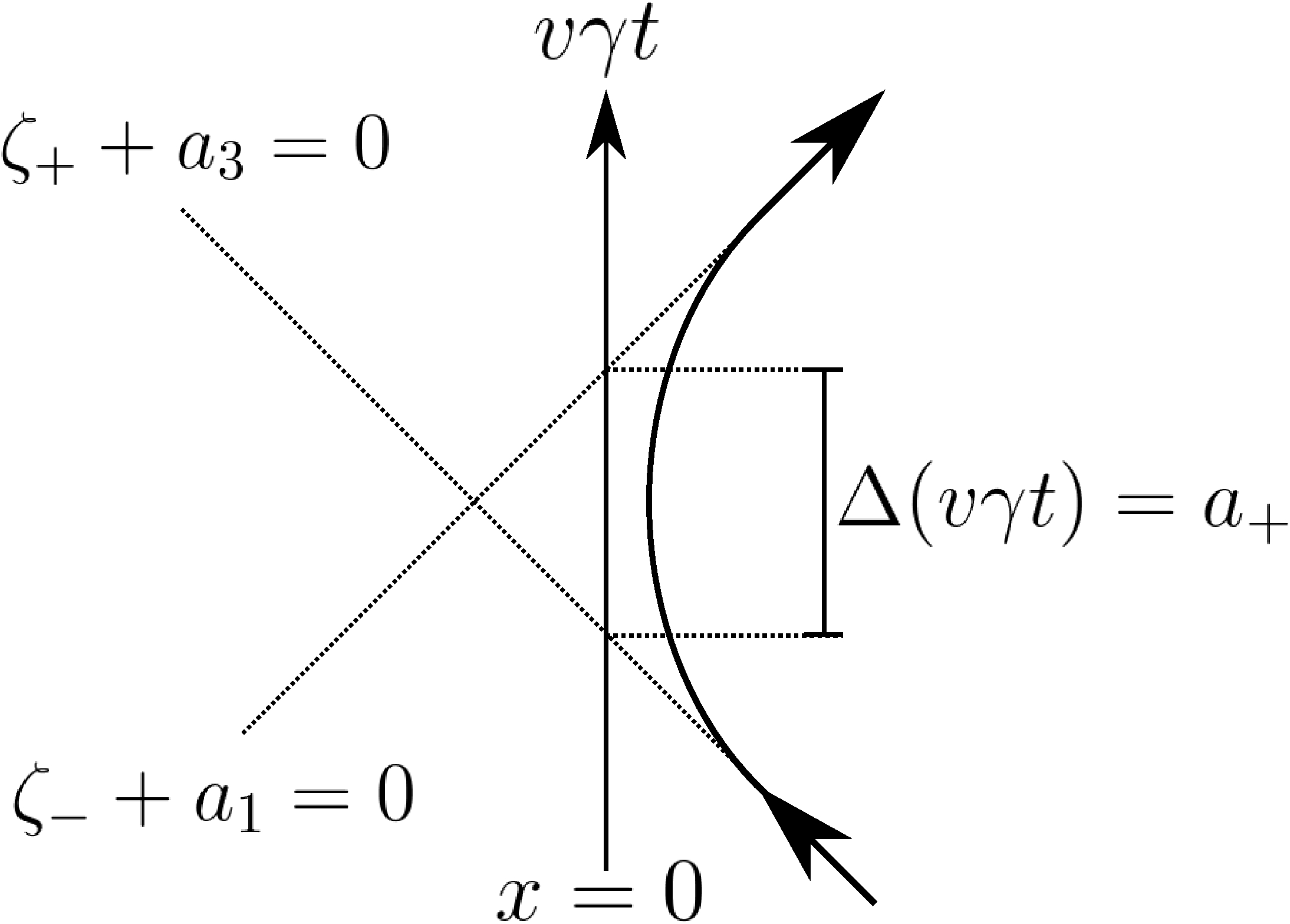}
\caption{The time shift due to boundary scattering via the method of images.\label{TimeDelay}}
\end{center}
\end{figure}

By comparing the times at which we expect the soliton centre to be at $x=0$, as extrapolated from its asymptotic in-going and out-going trajectories, we can deduce that the reflected soliton suffers a time shift (or phase shift) of $\Delta(v\gamma t)=a_{+}$.  This is depicted in Figure \ref{TimeDelay}.  The form of $a_{+}$ as a function of the soliton's velocity and the two boundary parameters can be calculated by substitution of the solution \eqref{MOISol} into the boundary condition \eqref{IBC}.  In terms of the parameterisation introduced by the authors of \cite{SSW},
\be
M{\rm cos}\left(\frac{\varphi_{0}}{2}\right)=2{\rm cosh}(\zeta){\rm cos}(\eta),\quad M{\rm sin}\left(\frac{\varphi_{0}}{2}\right)=2{\rm sinh}(\zeta){\rm sin}(\eta)\label{zetaeta}
\ee
where $0\leq\zeta<\infty$ and $-\pi<\eta\leq\pi$ the phase shift can be written as
\be
a_{+}={\rm ln}\left\{-\epsilon\kappa^{2}v^{2}\left[\frac{\left(\kappa^{2}+{\rm tan}^{2}\left(\frac{\eta}{2}\right)\right)\left(1-\kappa^{2}{\rm tanh}^{2}\left(\frac{\zeta}{2}\right)\right)}{\left(1+\kappa^{2}{\rm tan}^{2}\left(\frac{\eta}{2}\right)\right)\left(\kappa^{2}-{\rm tanh}^{2}\left(\frac{\zeta}{2}\right)\right)}\right]^{\pm 1}\right\}.\label{a}
\ee

\begin{figure}
\centering
\begin{tabular}{cc}
\includegraphics[width=0.35\textwidth]{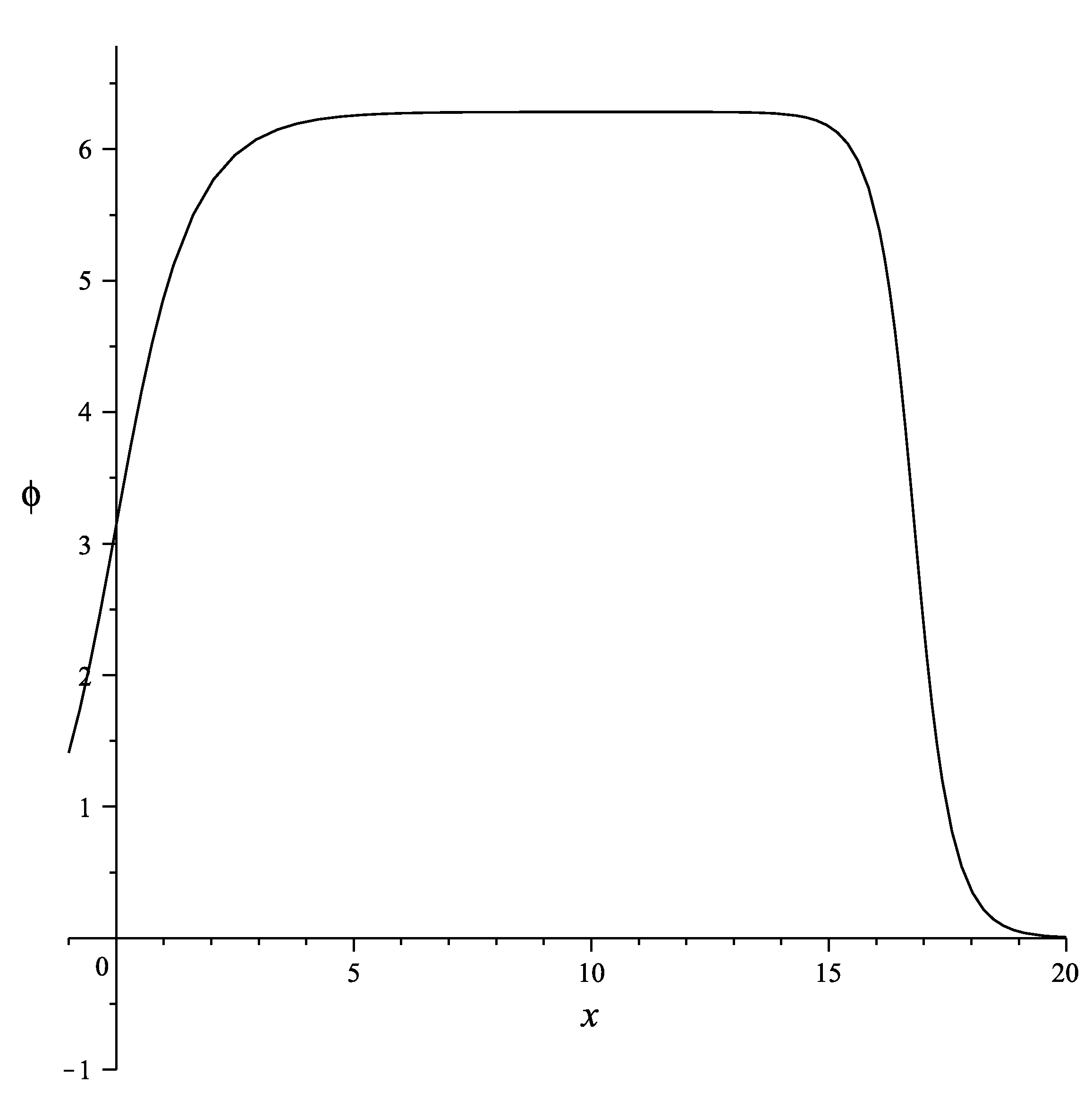} &
\includegraphics[width=0.35\textwidth]{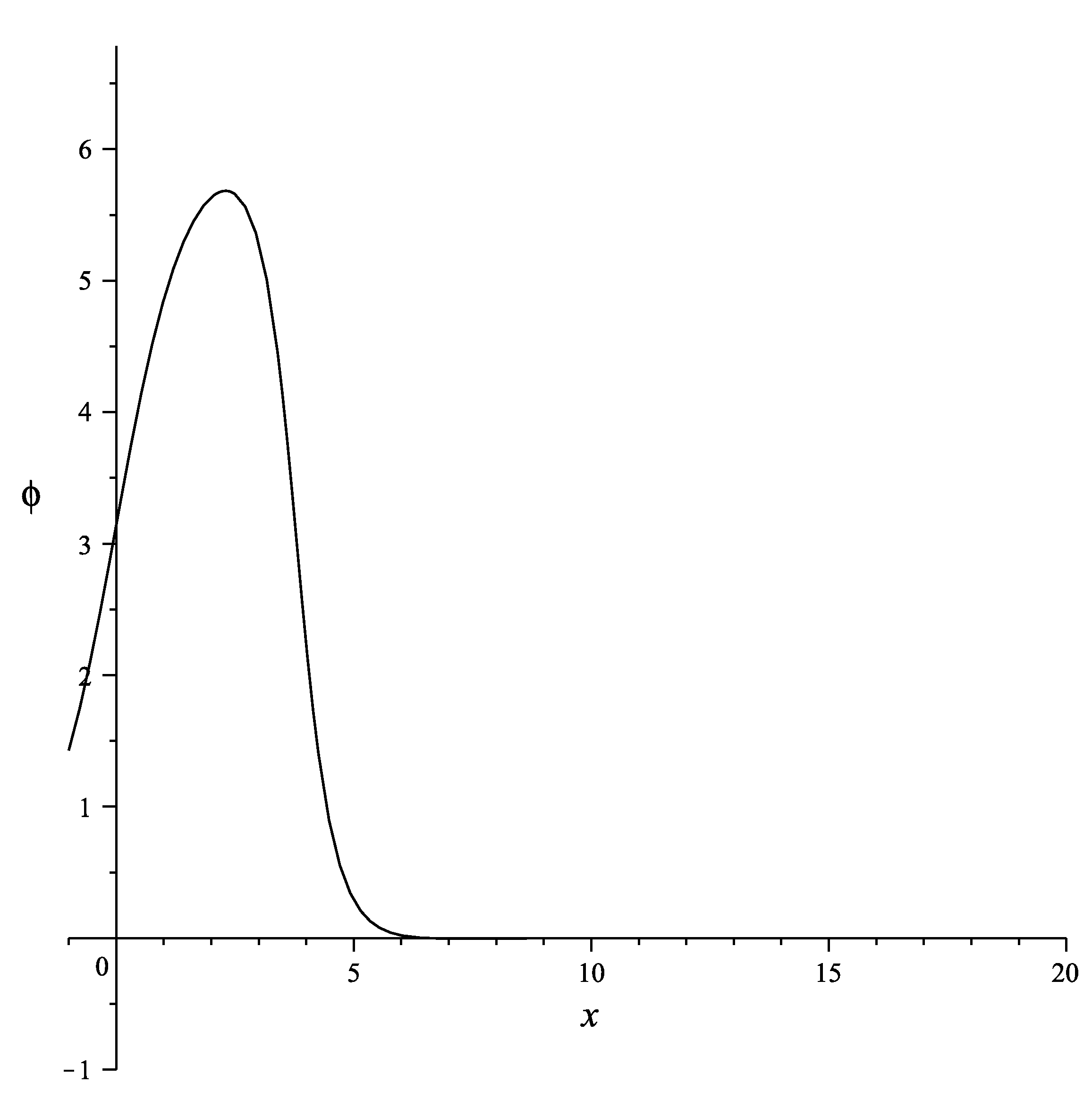} \\
\includegraphics[width=0.35\textwidth]{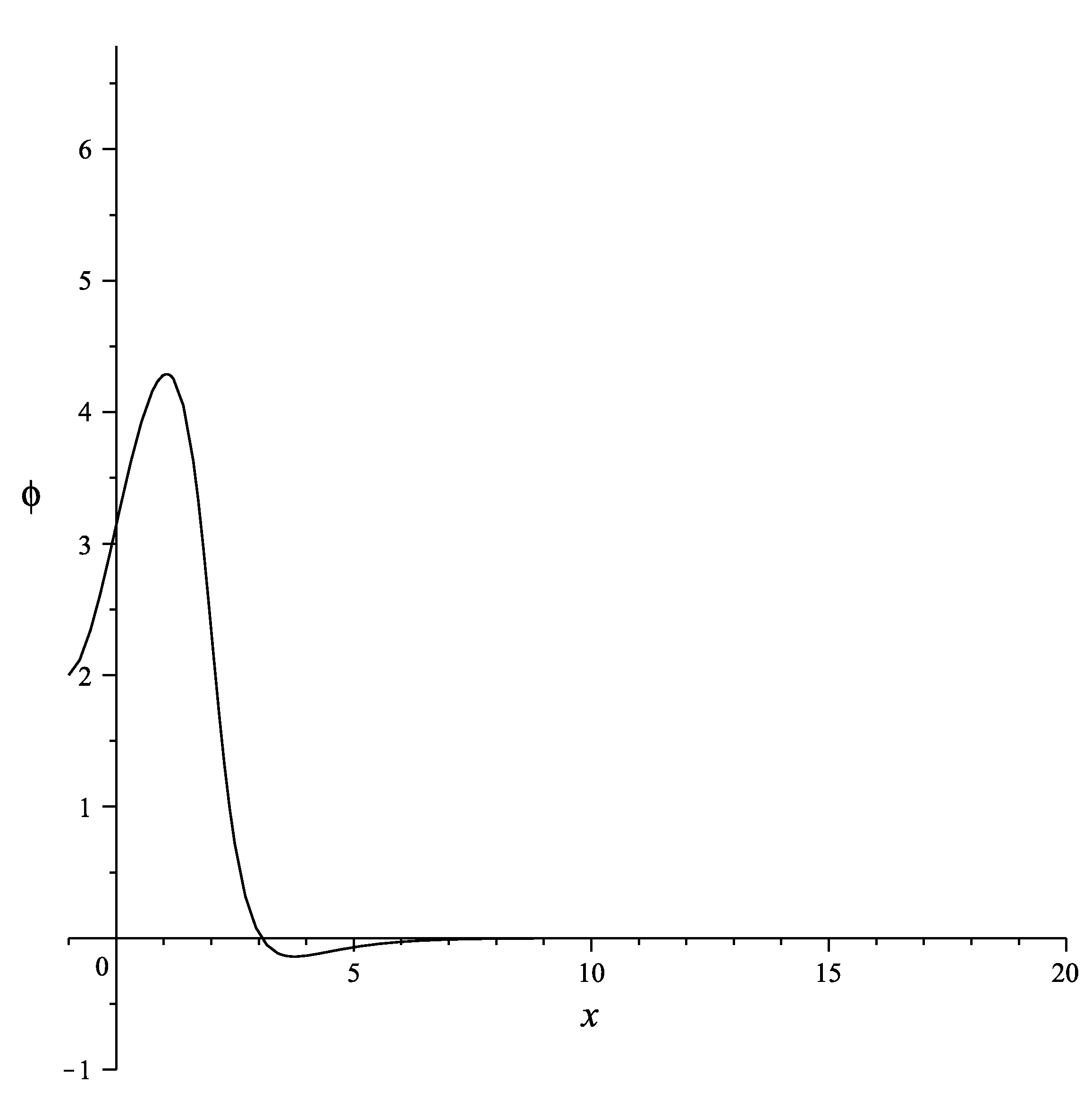} &
\includegraphics[width=0.35\textwidth]{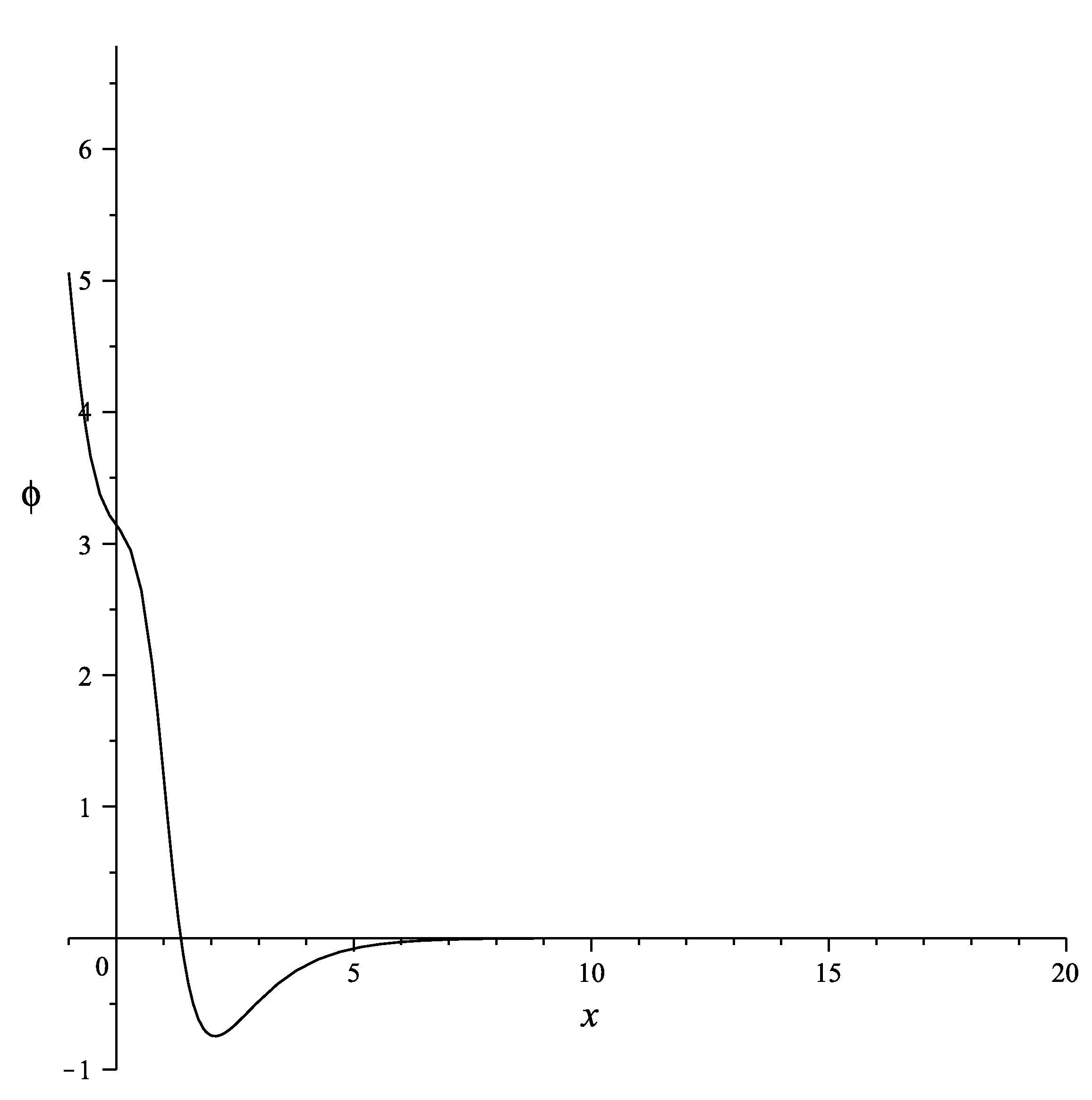} \\
\includegraphics[width=0.35\textwidth]{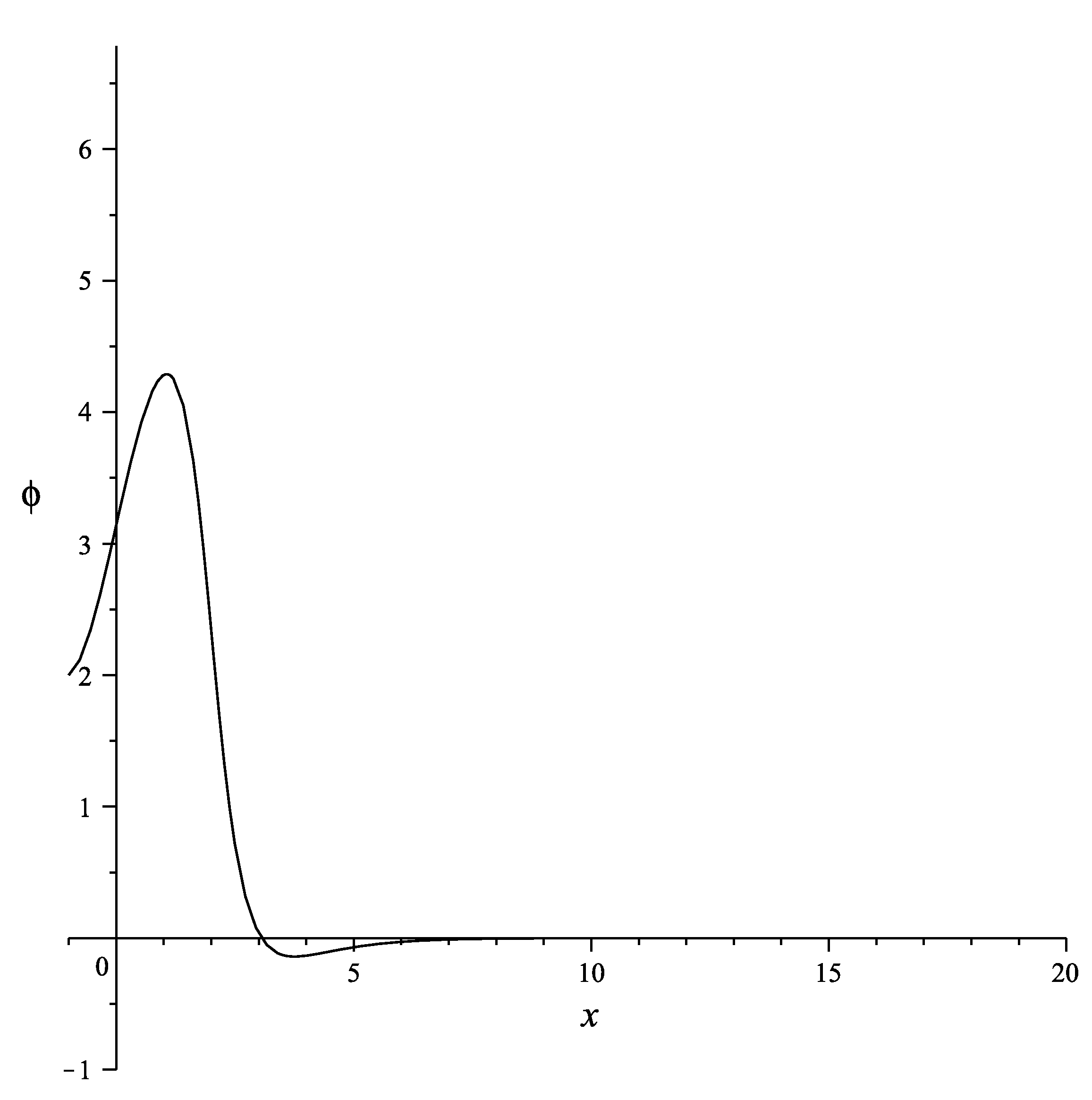} &
\includegraphics[width=0.35\textwidth]{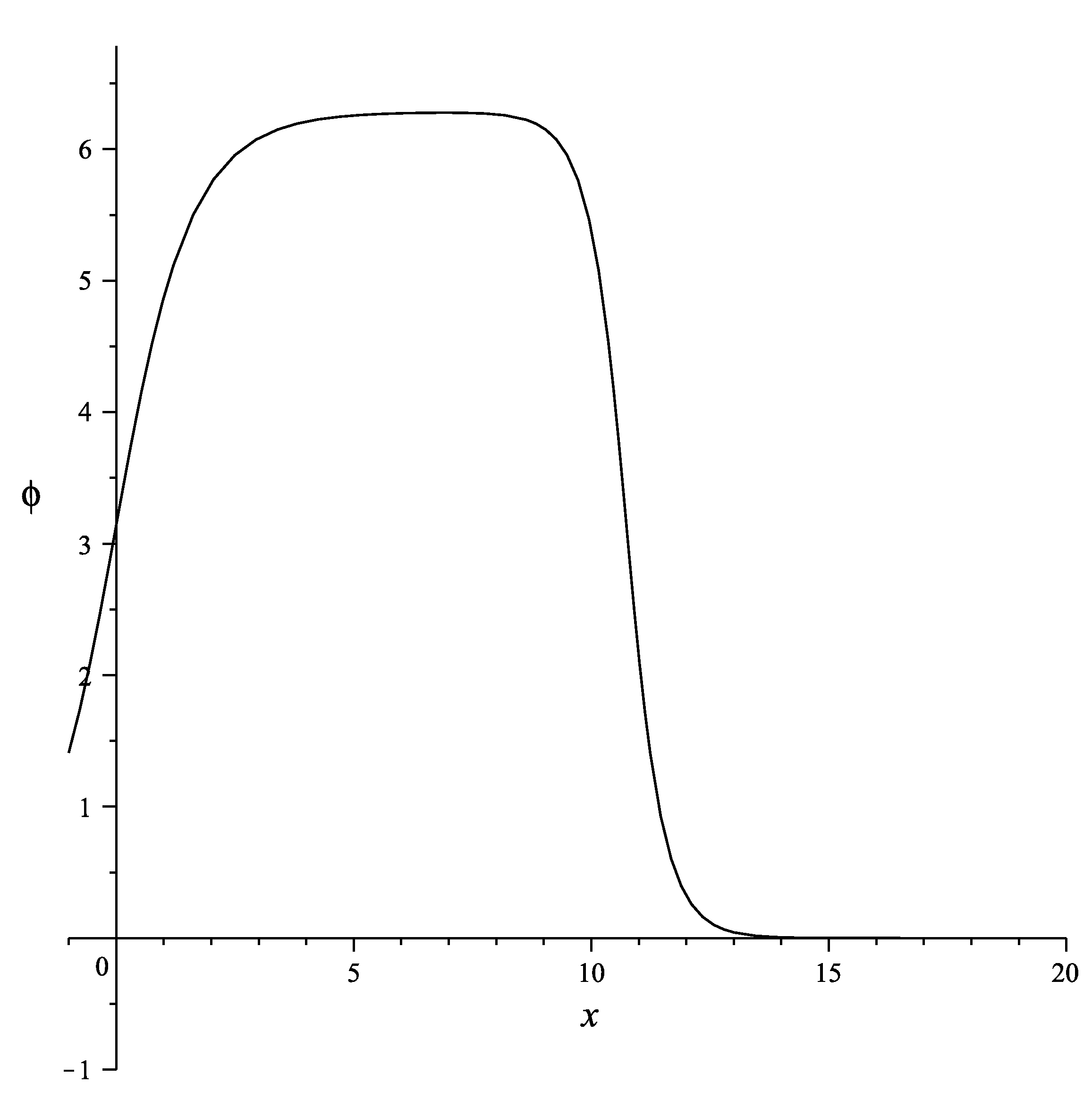}
\end{tabular}
\caption{An example of sine-Gordon boundary / soliton scattering using the method images.  Time flows from upper left to lower right.  In this instance a kink scatters with Dirichlet boundary conditions $\varphi_{0}=\pi$ at $x=0$ and $\varphi\to 0$ as $x\to \infty$.\label{SGSequence}}
\end{figure}

Given $a_{+}\in\mathbb{R}$ then we see that the argument of the log is positive as long as the combination $-\epsilon\left(\kappa^{2}-{\rm tanh}^{2}\left(\frac{\zeta}{2}\right)\right)$ is positive, according to which we must pick $\epsilon=\pm 1$.  In fact, for a given selection of boundary parameters $\{\zeta,\eta\}$ this leads to two distinct regimes of behaviour that depend upon the speed of the reflecting soliton.  Remembering that $\epsilon\equiv\epsilon_{1}\epsilon_{3}$ then the two regimes comprise Dirichlet-like solutions where solitons reflect as solitons and Neumann-like solutions where solitons reflect as anti-solitons:
\bea
|\theta|<\zeta\quad&\Ra&\quad \epsilon=+1,\quad {\rm Dirichlet~regime}\nn\\
|\theta|>\zeta\quad&\Ra&\quad \epsilon=-1,\quad {\rm Neumann~regime.}\nn
\eea
At the exact point between these two regimes, for $|\theta|=\zeta$, the phase delay diverges with the interpretation that the incoming soliton has been absorbed by the boundary.

When $M$ or, equivalently $\zeta$, is large then a greater range of $\theta$ falls into the Dirichlet regime so that the true Dirichlet condition is $M\to\infty$, which by \eqref{IBC} appropriately requires $\varphi|_{x=0}=\varphi_{0}$ if $\varphi'|_{x=0}$ is to be finite.  For the Dirichlet condition $a_{+}$ is given by
\be
a_{+}^{{\rm Dirichlet}}={\rm ln}\left\{(1-s)^{2}\left[\frac{1-s+(1+s){\rm tan}^{2}\left(\frac{\varphi_{0}}{4}\right)}{1+s+(1-s){\rm tan}^{2}\left(\frac{\varphi_{0}}{4}\right)}\right]^{\pm 1}\right\},\quad s\equiv\frac{1}{\gamma}.\label{aDirichlet}
\ee
$M=0$, or equivalently $\zeta=0$ and $\eta=\frac{\pi}{2}$ gives $\varphi'|_{x=0}=0$ and hence the true Neumann condition.

The choice in sign of the power in \eqref{a} reflects the choice in solutions we obtain for $a_{+}$, having solved a quadratic equation. The method of images in fact produces two valid integrable boundary solutions at the same time, one for $x\in (-\infty,0]$ and one for $x\in[0,\infty)$.  We may generally swap one for the other by an appropriate combination of operations $x\to-x$, $\varphi\to-\varphi$ and $\varphi\to\varphi\pm 2\pi$.

Figure \ref{SGSequence} plots an example of boundary / soliton scattering.

\section{Open string on $\mathbb{R}\times S^{2}$}

We will consider open strings obtained by carefully taking half of the $N=3$ magnon solution presented in \cite{KPV} in such a way that the motions of the string end point thus produced correspond to the integrable boundary of sine-Gordon theory, with the giant magnon reflecting off this end point being the reflecting sine-Gordon soliton.   We will need to include in the string solution the complex moduli discussed by the authors of \cite{KPV} but not explicitly considered.

The string solutions we describe live on an $S^{2}$ subspace of the $S^{5}$ of $AdS_{5}\times S^{5}$ so that we will not be concerned with the spacial coordinates of the $AdS_{5}$ part of the space; these are set to zero.  The target space time coordinate is provided by the global time coordinate of $AdS_{5}$ so that the strings move on $\mathbb{R}_{t}\times S^{2}$.  We set the common radius $R$ of $AdS_{5}$ and $S^{5}$ to be $R=1$.  As is often the case in the description of the giant magnon solutions we use the conformal and static partial gauges so that the world sheet time variable is aligned with the global time variable and the range of the world sheet spacial variable $x$ is infinite or semi-infinite due to the divergent energy of the string solutions.

Note that because the energy density of the string is constant in the conformal gauge and we take the solution supported on $x\in[0,\infty)$ we keep the length of the string in this gauge fixed and hence the energy of the open string thus defined will be constant.

In section 3.1 we will include and discuss the properties of the moduli of the bulk ($x\in(-\infty,\infty)$) solution restricted to $\mathbb{R}\times S^{2}$.  In section 3.2 we apply the method of images and analyse the asymptotics of the solution in order to find the dictionary between the open string and boundary sine-Gordon solutions that is essential to this work.

\subsection{N-magnon string solution on $\mathbb{R}\times S^{2}$ with moduli}

The solution presented by Kalousios, Papathanasiou and Volovich\cite{KPV} uses the dressing method\cite{Spradlin} to provide an $N$-magnon string solution that lives on $\mathbb{R}\times S^{3}$, determined by $N$ complex parameters $\lambda_{i}$.  If the coordinates on $S^{3}$ are $\{X_{3},X_{4},X_{5},X_{6}\}$ then for complexified coordinates $Z_{1}=X_{5}+iX_{6}$, $Z_{2}=X_{3}+iX_{4}$ it is
\bea
Z_{1}&=&\frac{e^{it}}{\prod^{N}_{l=1}|\lambda_{l}|}\frac{N_{1}}{D},\label{Z1}\\
Z_{2}&=&\frac{-ie^{-it}}{\prod^{N}_{l=1}|\lambda_{l}|}\frac{N_{2}}{D}\label{Z2}
\eea
with the indices $i,j$ running over the ordered values $1,\bar{1},2,\bar{2},...,N,\bar{N}$,
\bea
D&=&\displaystyle\sum_{\mu=0,1}\left(\prod^{2N}_{i<j}\lambda_{ij}[\mu_{i}\mu_{j}+(\mu_{i}-1)(\mu_{j}-1)]\right)\left(\displaystyle\prod^{2N}_{i=1}{\rm exp\left[\mu_{i}(2i\mathcal{Z}_{i})\right]}\right),\label{Dgeneral}\\
N_{1}&=&\displaystyle\sum_{\mu=0,1}\left(\prod^{2N}_{i<j}\lambda_{ij}[\mu_{i}\mu_{j}+(\mu_{i}-1)(\mu_{j}-1)]\right)\left(\displaystyle\prod^{2N}_{i=1}[\mu_{i}\lambda_{i}]{\rm exp\left[\mu_{i}(2i\mathcal{Z}_{i})\right]}\right),\label{N1general}\\
N_{2}&=&\displaystyle\sum_{\mu=0,1}\left(\prod^{2N}_{i<j}\lambda_{ij}[\mu_{i}\mu_{j}+(\mu_{i}-1)(\mu_{j}-1)]\right)\left(\displaystyle\prod^{2N}_{i=1}[-(\mu_{i}-1)\lambda_{i}]{\rm exp\left[\mu_{i}(2i\mathcal{Z}_{i})\right]}\right)\nn\\\label{N2general}
\eea
where $\lambda_{ij}\equiv \lambda_{i}-\lambda_{j}$.  The world sheet coordinates enter through
\be
\mathcal{Z}_{i}=\frac{z}{\lambda_{i}-1}+\frac{\bar{z}}{\lambda_{i}+1}\label{Zi}
\ee
in which $z=\frac{1}{2}(x-t)$ and $\bar{z}=\frac{1}{2}(x+t)$ are the light-cone coordinates and the sums over $\mu_{i}=0,1$ are subject to the condition that
\be
\displaystyle\sum_{i=1}^{2N}=\bigg\{
\begin{array}{ccc}
N, &{\rm for }&N_{1}, D\\
N+1,&{\rm for }&N_{2}.
\end{array}
\ee
The parameters $\lambda_{k}$ can be written $\lambda_{k}=r_{k}e^{i\frac{p_{k}}{2}}$ where $p_{k}$ is the magnon momentum.  If we take $|r_{k}|=1$ then the solution reduces to the solution on $\mathbb{R}\times S^{2}$ as $Z_{2}$ becomes, up to a factor of $i e^{4it}$ that cancels, pure real.  Writing also
\be
v_{k}={\rm cos}\left(\frac{p_{k}}{2}\right),\quad \gamma_{k}=\frac{1}{{\rm sin}\left(\frac{p_{k}}{2}\right)}\label{vgamma}
\ee
the $\mathcal{Z}_{k}$ defined in \eqref{Zi} then become

\bea
\mathcal{Z}_{k}&=&\frac{-i\gamma_{k}}{2}x+\frac{i\gamma_{k}v_{k}}{2}+\frac{t}{2}\nonumber\\
&=&\frac{-i}{2}\zeta_{k}+\frac{t}{2}
\eea
and then
\be
e^{2i\mathcal{Z}_{k}}=e^{\zeta_{k}+it},\qquad \zeta_{k}\equiv \gamma_{k}(x-v_{k}t).\label{Zzeta}
\ee
The $v_{k}$ are then the velocities of the solitons moving on the world sheet, equivalently the velocities of the sine-Gordon kinks and anti-kinks to which they correspond.
\newline

The finite difference between the string energy $E\equiv \Delta$ and angular momentum $J$ that characterise the giant magnon solutions is here the sum of the values for $N$ individual giant magnons,
\be
\Delta-J=\frac{\sqrt{\lambda}}{\pi}\sum_{k=1}^{N}{\rm sin}\frac{p_{k}}{2}
\ee
and the solutions will satisfy both of the Virasoro constraints
\bea
\dot{X}_{i}\dot{X}_{i}+X'_{i}X'_{i}&=&1,\qquad i=1,2,3\label{V1}\\
\dot{X}_{i}X'_{i}&=&0.\label{V2}
\eea
\nl

As discussed in \cite{KPV,Spradlin} the dressing method admits a generalisation of the solution whereby
\be
e^{i\mathcal{Z}_{k}}\to w~e^{i\mathcal{Z}_{k}},\qquad w\in\mathbb{C}
\ee
introducing an extra $2N$ real parametric degrees of freedom, two of which may always be absorbed by shifts of $x$ and $t$, thus
\be
\mathcal{Z}_{k}\to\mathcal{Z}_{k}+i{\rm ln}(w_{k}).\label{Zshift}
\ee
This leaves $2(N-1)$ parameters to be moduli of the solution.  Clearly the single magnon solution has no moduli.  A 3-magnon solution would have four moduli, but by breaking spacial translational symmetry we shall be left a fifth.  However, we will require that the solution be constrained from $\mathbb{R}\times S^{3}$ to $\mathbb{R}\times S^{2}$ so that three moduli become discretised leaving us two continuous moduli that will correspond to the two boundary parameters encountered in the sine-Gordon picture and three discrete soliton/anti-soliton parameters.

Including these parameters then as in \eqref{Zshift} with $w_{k}={\rm exp}\left(\frac{\alpha_{k}}{2}+i\psi\right)$, we have
\be
e^{2i\mathcal{Z}_{k}}\to \epsilon_{k}e^{\zeta_{k}-\alpha_{k}+it},\quad \alpha_{k}\in\mathbb{R}\label{expfactors}
\ee
where we used \eqref{Zzeta} and defined
\be
\epsilon_{k}\equiv e^{-2i\psi},\quad \psi\in\mathbb{R}.
\ee
The solution is then composed, via \eqref{Z1} to \eqref{N2general}, of sums of products of factors \eqref{expfactors} together with coefficients involving functions of the magnon momenta $p_{k}$.  In this raw form, such as is included explicitly in an appendix in \cite{KPV}, the 3-magnon solution is difficult to analyse and so we first reduce it to products of trigonometric and hyper trigonometric functions.  The general 3-magnon solution in such a form is included here in appendix A.  The form will further simplify with our specific application.

If we examine the solution in Appendix A it is clear that the factors $\epsilon_{k}$ which appear variously in products of $\epsilon_{1}\epsilon_{2}\epsilon_{3}$, $\epsilon_{i}\epsilon_{j}$ and on their own, must be equal to $\pm 1$ in order that $Z_{2}$ be real.  Three of the six moduli we have introduced therefore become discretised in order that we constrain the solution to the 2-sphere:
\be
\epsilon_{k}=\pm 1;\qquad \psi=m\frac{\pi}{2},~~m\in\mathbb{Z}.
\ee
This recovers the soliton/anti-soliton parameters we met in the sine-Gordon picture and in fact reiterates the appearance of the topological solitons of sine-Gordon theory from the non-topological solitons of the complex sine-Gordon theory\cite{Lund:1976ze} to which the string theory on $\mathbb{R}\times S^{3}$ corresponds.
\newline\newline
\indent Before we turn to the method of images we pause to comment on the generic behaviour of the $N=3$ solutions on $\mathbb{R}\times S^{2}$.

We recall that at early and late times giant magnons form semi-circular arcs with their end points ($x\to\pm\infty$) on the equator which is defined by the axis about which the string has a divergent angular momentum.  The radius of this arc is equal to ${\rm sin}\left(\frac{p}{2}\right)$ in terms of the world sheet momentum that creates the excitation, or $\sqrt{1-v^{2}}$ in terms of its speed on the world sheet or correspondingly of the sine-Gordon soliton.  Fast solitons are therefore small, vanishing to a point on the equator at $v=1$, while a stationary soliton crosses the pole of the sphere as half of a great circle.  

For a single giant magnon the physics is invariant under $Z_{2}\to-Z_{2}$, as is the physics of a single sine-Gordon kink or anti-kink.  For multi-magnon states though we may choose which hemisphere each magnon lives upon, just as there is a difference between multi-soliton sine-Gordon solutions in the choice between composing kinks with kinks or anti-kinks.  While the phase delays of scattering kinks and anti-kinks, and hence the scattering time delays of giant magnons with those on the same or opposite hemispheres, are identical, the solutions in each case produce qualitatively different behaviour.

\begin{figure}
\centering
\includegraphics[width=0.4\textwidth]{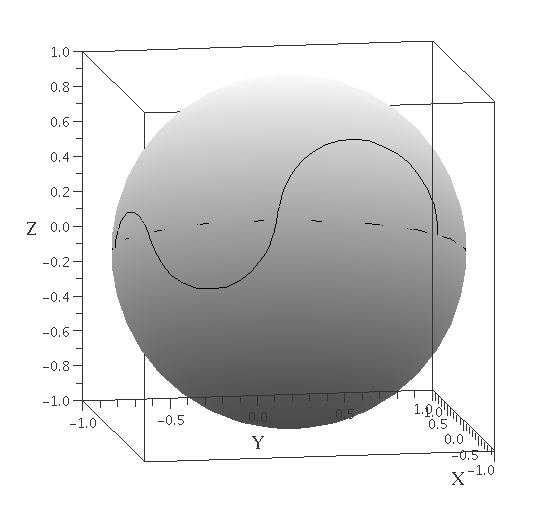}
\caption{A 3-magnon solution at early time.  The momenta are $p_{1}=\frac{\pi}{6}$, $p_{2}=\frac{\pi}{4}$ and $p_{3}=\frac{\pi}{3}$. The moduli chosen are $\epsilon_{k}=+1$ and $\alpha_{k}=0$, $k=1,2,3$, and the corresponding sine-Gordon solution is kink/anti-kink/kink.\label{3MagEarly}}
\end{figure} 

\begin{figure}
\centering
\includegraphics[width=0.4\textwidth]{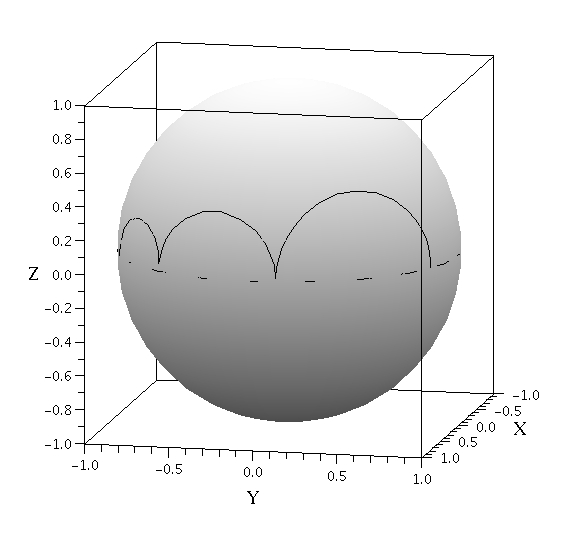}
\caption{Another 3-magnon solution at early time.  The momenta are again $p_{1}=\frac{\pi}{6}$, $p_{2}=\frac{\pi}{4}$ and $p_{3}=\frac{\pi}{3}$, but this time the moduli chosen are $\epsilon_{1}=\epsilon_{3}=+1$ and $\epsilon_{2}=-1$. $\alpha_{k}=0$, $k=1,2,3$ again and the corresponding sine-Gordon solution is kink/kink/kink.\label{3MagEarly_B}}
\end{figure}

The solution as presented in \cite{KPV} effectively has $\epsilon_{1}=\epsilon_{2}=\epsilon_{3}=1$ and $\alpha_{1}=\alpha_{2}=\alpha_{3}=0$, an example of which is plotted in Figure \ref{3MagEarly} at an early time.  We may think that due to all $\epsilon_{k}=+1$ then the corresponding sine-Gordon solution is a 3-kink, it is however a kink/anti-kink/kink solution, as ordered at $t\to-\infty$.  In fact, when the momenta are ordered such as $p_{1}<p_{2}<p_{3}$ then $\epsilon_{2}$ in the string picture is always negative the similar parameter coming from the sine-Gordon picture, and the `raw' 3-magnon solution of \cite{KPV}, when restricted to the 2-sphere by taking all $\lambda_{k}$ to have $|\lambda_{k}|=1$, will always give alternating kink/anti-kink/kink or anti-kink/kink/anti-kink solutions.  Had we therefore chosen $\epsilon_{2}=-1$ we would obtain a 3-magnon solution where each giant magnon sits on the same hemisphere and the corresponding sine-Gordon solution is kink/kink/kink.  Such a choice is plotted in Figure \ref{3MagEarly_B}.

Each of the inequivalent choices of $\{\epsilon_{k}\}$ (those not related by each $\epsilon_{k}\to-\epsilon_{k}$, hence the coordinate $Z_{2}\to-Z_{2}$) display different finite time behaviour as the solitons scatter on the world sheet, but the qualitative difference is between those solutions with sine-Gordon topological charge $|Q|=1$ (such as kink/anti-kink/anti-kink) and those with $|Q|=3$ (such as kink/kink/kink).

The $|Q|=3$ solutions display cusps on the bulk of the string that persist at finite time with the result that the scattering appears as three giant magnons that smoothly grow or shrink such that their orders are reversed as we go from $t\to-\infty$ to $t\to+\infty$.  On the other hand those solutions with $|Q|=1$ display cusps that emerge from the equatorial end points of the string and race through to the opposite end point where they merge, sometimes annihilating with another cusp along the way before reappearing.

This behaviour of the cusps is given in the sine-Gordon theory by the behaviour of points where $\varphi(x,t)=2m\pi,~m\in\mathbb{Z}$, as is seen from the relation
\be
{\rm sin}^{2}\frac{\varphi}{2}=X_{i}'X_{i}'.
\ee
Taking the kink / boundary scattering plotted in Figure \ref{SGSequence} (Section 2) for instance, it may appear at first glance that $\varphi\to 0$ from above as $x\to \infty$, especially at early and late times.  However, $\varphi\to 0$ from \emph{below} $\forall~t$ as $x\to \infty$ so that we must always have a point where $\varphi=0$, and hence a cusp, $X_{i}'=0$.

\subsection{Method of images on $\mathbb{R}\times S^{2}$}

In this subsection we will select the 3-magnon solution required by the method of images and find the string moduli in terms of the sine-Gordon boundary parameters so that we can say exactly which sine-Gordon integrable boundary configuration corresponds to which open string.  

We proceed with the method of images then as in Section 2 by taking
\bea
v_{1}\equiv v,&\quad v_{2}=0,&\quad v_{3}=-v\\
\gamma_{1}\equiv\gamma,&\quad\gamma_{2}=1,&\quad\gamma_{3}=\gamma
\eea
or in terms of the magnon momenta, using \eqref{vgamma}, and the abbreviations $s_{k}\equiv {\rm sin}\left(\frac{p_{k}}{2}\right)$, $c_{k}\equiv {\rm cos}\left(\frac{p_{k}}{2}\right)$,
\bea
c_{1}\equiv c,&\quad c_{2}=0,&\quad c_{3}=-c\\
s_{1}\equiv s,&\quad s_{2}=1,&\quad s_{3}=s.
\eea
Note that $v\to -v$ is equivalent to $p\to 2\pi-p$. Mirroring the notation adopted in the previous section we write
\be
\alpha_{+}\equiv \alpha_{1}+\alpha_{3},\quad \alpha_{-}=\alpha_{1}-\alpha_{3},\quad\alpha_{2}\equiv\beta
\ee
and we note that in precisely the manner in which we were able to perform an overall time shift to absorb $a_{-}$ from the sine-Gordon solution, so we are able to absorb $\alpha_{-}$ so that it does not appear.

Taking the general solution presented in appendix A and inserting the above we obtain forms of $D$, $N_{1}$ and $N_{2}$ from which factors of $32c^{2}e^{3it}$ will cancel.  A further factor of $ie^{it}$ cancels from $N_{2}$.  We produce below these simplified forms $\tilde{D}$, $\tilde{N}_{1}$ and $\tilde{N}_{2}$ which are such that the ratios
\be
\tilde{Z}_{1}=\frac{\tilde{N}_{1}}{\tilde{D}}\quad {\rm and}\quad \tilde{Z}_{2}=\frac{\tilde{N_{2}}}{\tilde{D}}\label{ZA}
\ee
give the string solution in a co-moving frame.  In the solution below upper and lower signs correlate with the upper and lower cases and correspond to taking $\epsilon=\pm 1$.  Similarly to the previous section we have used
\be
\epsilon=\epsilon_{1}\epsilon_{3},\qquad \epsilon_{3}=\epsilon\epsilon_{1}
\ee
which has been used to eliminate $\epsilon_{3}$ in favour of $\epsilon$.

\bea
\tilde{D}=2s^{2}\epsilon_{0}{\rm cosh}(x-\beta)+8s\epsilon_{1}\left\{\begin{array}{c}-{\rm cosh}(v\gamma t){\rm cosh}(\gamma x-\frac{\alpha_{+}}{2})\\+{\rm sinh}(v\gamma t){\rm sinh}(\gamma x-\frac{\alpha_{+}}{2})\end{array}\nn\right\}\hspace{3cm}\nn\\
\pm\epsilon_{0}\left[(1-s)^{2}{\rm cosh}((2\gamma+1)x-\alpha_{+}-\beta)+(1+s)^{2}{\rm cosh}((2\gamma-1)x-\alpha_{+}+\beta)\right.\nn\\
\left.+2{\rm cosh}(2v\gamma t)\right],\hspace{3cm}\label{DA}
\eea

\bea
\tilde{N}_{1}=8sc\epsilon_{1}\left\{\begin{array}{c}+{\rm sinh}(v\gamma t){\rm sinh}(\gamma x-\frac{\alpha_{+}}{2})\\-{\rm cosh}(v\gamma t){\rm cosh}(\gamma x-\frac{\alpha_{+}}{2})\end{array}\right\} \mp 4sc\epsilon_{0}{\rm sinh}(x-\beta){\rm sinh}(2v\gamma t)\hspace{1cm}\nn\\\nn\\
\left.+i\left[2s^{2}\epsilon_{0}{\rm sinh}(x-\beta)+8s^{2}\epsilon_{1}\left\{\begin{array}{c}-{\rm cosh}(v\gamma t){\rm sinh}(\gamma x-\frac{\alpha_{+}}{2})\\+{\rm sinh}(v\gamma t){\rm cosh}(\gamma x-\frac{\alpha_{+}}{2})\end{array}\right\}\right.\right.\nn\hspace{1cm}\\
\left.\pm\epsilon_{0}\left[-(1-s)^{2}{\rm sinh}((2\gamma+1)x-\alpha_{+}-\beta)\right.\right.\nn\hspace{3cm}\\
\left.\left.+(1+s)^{2}{\rm sinh}((2\gamma-1)x-\alpha_{+}+\beta)+2(1-2c^{2}){\rm sinh}(x-\beta){\rm cosh}(2v\gamma t)\right]\right.\Big],\hspace{0.2cm}\label{N1A}
\eea

\bea
\tilde{N}_{2}=-6s^{2}+4s(1-s)\epsilon_{0}\epsilon_{1}\left\{\begin{array}{c}+{\rm cosh}(v\gamma t){\rm cosh}((1+\gamma)x-\beta-\frac{\alpha_{+}}{2})\nn\\-{\rm sinh}(v\gamma t){\rm sinh}((1+\gamma)x-\beta-\frac{\alpha_{+}}{2})\end{array}\right\}\hspace{1.5cm}\nn\\
+4s(1+s)\epsilon_{0}\epsilon_{1}\left\{\begin{array}{c}+{\rm cosh}(v\gamma t){\rm cosh}((1-\gamma)x-\beta+\frac{\alpha_{+}}{2})\nn\\+{\rm sinh}(v\gamma t){\rm sinh}((1-\gamma)x-\beta+\frac{\alpha_{+}}{2})\end{array}\right\}\hspace{1.5cm}\nn\\
\mp 2\left[c^{2}{\rm cosh}(2\gamma x-\alpha_{+})+{\rm cosh}(2v\gamma t)\right].\hspace{3.4cm}\label{N2A}
\eea
\newline
\indent Our job is to make identifications between the displacement parameters on both sides which we achieve by examination of the asymptotics.  We define
\be
\zeta_{1}\equiv\zeta_{-}\equiv\gamma(x-vt),\quad\zeta_{3}\equiv\zeta_{+}\equiv\gamma(x+vt)
\ee
and using equations \eqref{ZA} and \eqref{DA} to \eqref{N2A} (with $v\gamma t\to v\gamma t +\alpha_{-}$) we examine the asymptotic form of the right-mover with target space coordinates $(X_{-},Y_{-},Z_{-})$ to find
\bea
X_{-}&\to&\frac{-4sc~{\rm exp}({-\zeta_{-}-{\rm ln}(1-s)+\alpha_{1}})}{{\rm cosh}(\zeta_{-}+{\rm ln}(1-s)-\alpha_{1})}\\
Y_{-}&\to&\frac{-2{\rm exp}(\zeta_{-}+{\rm ln}(1-s)-\alpha_{1})+2(1-2c^{2}){\rm exp}(\zeta_{-}+{\rm ln}(1-s)-\alpha_{1})}{{\rm cosh}(\zeta_{-}+{\rm ln}(1-s)-\alpha_{1})}\hspace{0.8cm}\\
Z_{-}&\to&\frac{s\epsilon_{1}}{{\rm cosh}(\zeta_{-}+{\rm ln}(1-s)-\alpha_{1})}.
\eea
As already proved by the authors of \cite{KPV} this should amount to the simple solution for a single giant magnon rotated in the $Z_{1}$ plane by an angle $\frac{p}{2}$ from parallel with the $X_{5}$-axis.  Performing the rotation thus
\be
\left(\begin{array}{c}\tilde{X}_{-}\\\tilde{Y}_{-}\end{array}\right)=\left(\begin{array}{cc}c&-s\\s&c\end{array}\right)\left(\begin{array}{c}X_{-}\\Y_{-}\end{array}\right)
\ee
we do indeed obtain the original giant magnon solution in co-moving coordinates,
\bea
\tilde{X}_{-}+i\tilde{Y}_{-}&=&{\rm sin}\left(\frac{p}{2}\right)~{\rm tanh}(\zeta_{-}+{\rm ln}(1-s)-\alpha_{1})-i{\rm cos}\left(\frac{p}{2}\right)\nn\\
Z_{-}&=&\frac{\epsilon_{1}s}{{\rm cosh}(\zeta_{-}+{\rm ln}(1-s)-\alpha_{1})},
\eea
together with a phase shift dependent upon $p$ and, of course, $\alpha_{1}$.  Dealing similarly with the left-mover with coordinates $(X_{+},Y_{+},Z_{+})$, rotating in the opposite direction as
\be
\left(\begin{array}{c}\tilde{X}_{+}\\\tilde{Y}_{+}\end{array}\right)=\left(\begin{array}{cc}c&s\\-s&c\end{array}\right)\left(\begin{array}{c}X_{+}\\Y_{+}\end{array}\right)
\ee
we obtain
\bea
\tilde{X}_{+}+i\tilde{Y}_{+}&=&-{\rm sin}\left(\frac{p}{2}\right){\rm tanh}(\zeta_{+}+{\rm ln}(1-s)-\alpha_{3})-i{\rm cos}\left(\frac{p}{2}\right)\nn\\
Z_{+}&=&\frac{\epsilon_{3}s}{{\rm cosh}(\zeta_{+}+{\rm ln}(1-s)-\alpha_{3})},
\eea
showing that the early and late time configurations are reflections of one another in the $Y$-axis of the co-moving frame, together with a potential mirroring in $Z_{2}$ if $\epsilon_{3}=-\epsilon_{1}$.

Taking $t\to\pm\infty$ with $x$ fixed we find the central soliton,
\bea
t\to-\infty,&\quad (s~{\rm tanh}(x-\beta),-c~{\rm tanh}(x-\beta),-\epsilon_{0}{\rm sech}(x-\beta))\label{CSminus}\\
t\to+\infty,&\quad (-s~{\rm tanh}(x-\beta),-c~{\rm tanh}(x-\beta),-\epsilon_{0}{\rm sech}(x-\beta)).\label{CSplus}
\eea
The corresponding asymptotic sine-Gordon solutions are easily found once we `spin up' these solutions by applying $\tilde{Z}_{1}\to e^{it}\tilde{Z}_{1}=Z_{1}$.  Comparing with the asymptotics obtained in the previous section,
\be
{\rm sin}\left(\frac{\varphi}{2}\right)=\frac{\epsilon_{3}}{{\rm cosh}(\zeta_{+}+a_{3})}=\frac{\pm 1}{{\rm cosh}(\zeta_{+}+{\rm ln}(1-s)-\alpha_{3})}
\ee
and
\be
{\rm sin}\left(\frac{\varphi}{2}\right)=\frac{\epsilon_{1}}{{\rm cosh}(\zeta_{-}+a_{1})}=\frac{\pm 1}{{\rm cosh}(\zeta_{-}+{\rm ln}(1-s)-\alpha_{1})}
\ee
giving
\be
a_{1}={\rm ln}(1-s)-\alpha_{1},{\rm and}\quad a_{3}={\rm ln}(1-s)-\alpha_{3},
\ee
or
\be
a_{+}=2{\rm ln}(1-s)-\alpha_{+}\label{aplus}
\ee
and for the central soliton
\be
{\rm sin}\left(\frac{\varphi}{2}\right)=\frac{-\epsilon_{0}}{{\rm cosh}(x+b-2{\rm ln}(\kappa))}=\frac{\pm 1}{{\rm cosh}(x-\beta)}
\ee
so
\be
b={\rm ln}\left(\frac{1-s}{1+s}\right)-\beta.\label{b}
\ee
From \eqref{bphi0} then we can related $\beta$ to $\varphi_{0}$ as
\be
\beta=-{\rm ln}\left(-\epsilon_{0}{\rm tan}\left(\frac{\varphi_{0}}{4}\right)\right).\label{betaphi0}
\ee

Relations \eqref{aplus} and \eqref{b}, together with the identifications
\be
\epsilon_{1}^{{SG}}=\epsilon_{1}^{{\rm String}},~~\epsilon_{3}^{{SG}}=\epsilon_{3}^{{\rm String}},~~-\epsilon_{0}=\epsilon_{2}^{{\rm String}},
\ee
give us the dictionary by which we can match sine-Gordon boundary solutions to corresponding string solutions through the two real moduli that appear in each theory plus boundary conditions at infinity.

\section{Scattering giant magnons off giant gravitons and semi-classical phase shifts}

In this section we will consider explicit examples of giant magnons in scattering with giant gravitons, beginning with three maximal cases and then turning to the case of non-maximal ($Y=0$) giant gravitons.  In each case we look for boundary parameters that produce the required string end point motions.

The coordinates of the embedding of $S^{5}$ and $S^{2}$ into $\mathbb{R}^{6}$ shall be given by six real coordinates $\{X_{i}\},~i=1,..,6$ or three complex coordinates
\be
W=X_{1}+iX_{2},\quad Y=X_{3}+iX_{4},\quad Z=X_{5}+iX_{6}
\ee
always satisfying
\be
|W|^{2}+|Y|^{2}+|Z|^{2}=1.\label{S5}
\ee
The giant gravitons wrap an $S^{3}\subset S^{5}$ so that their orientations are specified by any two relations $a_{i}X_{i}=b_{i}X_{i}=0$ for arbitrary vectors $a_{i},b_{i}\in\mathbb{R}^{6}$.  We follow Hofman and Maldacena's convention\cite{HofmanII} of taking the large angular momentum possessed by the strings and branes to be in the $\{X_{5},X_{6}\}$ direction and hence referring to the physically distinct orientations of the branes as either $Y=0$ or $Z=0$.

We shall use the semi-classical Levinson's theorem\cite{Jackiw:1975im} to calculate the leading contribution to the strong coupling scattering phase $e^{i\delta}$, given our knowledge of the time shift,
\be
\Delta T_{B}=\frac{\partial \delta_{B}}{\partial E},
\ee
due to integrable boundary scattering of sine-Gordon solitons of the Pohlmeyer reduced string.  Collisions with the boundary are elastic, so the speeds of the incoming and outgoing solitons are equal, and so it makes sense to identify the time delay involved in scattering as the sine-Gordon phase delay $a(p)$ divided by $v\gamma$,
\bea
\Delta T_{B}&=&\frac{a}{v\gamma},\qquad \gamma=\frac{1}{\sqrt{1-v^{2}}}\\
&=&\frac{{\rm sin}\frac{p}{2}}{{\rm cos}\frac{p}{2}}a(p).
\eea
We then obtain the \emph{magnon} phase shift $\delta$ by integrating with respect to the \emph{magnon} energy, remembering that, as discussed in \cite{Hofman:2006xt}, while the string and sine-Gordon pictures share their time coordinate exactly the energy of the string and sine-Gordon solitons differ.  Hence 
\be
\delta^{{\rm magnon}}=\int\frac{{\rm sin}\frac{p}{2}}{{\rm cos}\frac{p}{2}}a(p){\rm d}E(p)^{{\rm magnon}}.\label{deltaB}
\ee
\subsection{Scattering with maximal $Z=0$ and $Y=0$ giant gravitons}

In this subsection we consider the moduli necessary to construct the solutions to three previously studied instances of giant magnons scattering with maximal giant gravitons.
\vspace{0.5cm}

\noindent{\bf Maximal Y=0 cases} 
\nl\nl
The scattering of giant magnons with a maximal $Y=0$ giant graviton was studied in \cite{HofmanII} where the it was shown that there are two distinct scenarios dependent upon the state of the magnon, or equivalently the relative orientation of the giant magnon with respect to the brane.  As the string rotates in $Z=X_{5}+iX_{6}$ and lives on an $S^{2}$ this relative orientation is decided by the choice of the string's third coordinate.  The $Y=0$ brane coordinates satisfy
\be
X_{1}^{2}+X_{2}^{2}+X_{5}^{2}+X_{6}^{2}=1
\ee
so that the intersection of this brane with a maximal radius $S^{2}$ upon which the string moves is determined by whether we pick one of the $Y$ coordinates ($X_{3}$ or $X_{4}$) or one of the $W$ coordinates ($X_{1}$ or $X_{2}$) as the string's third coordinate.  Choosing the string's $S^{2}$ to be embedded in $\{X_{3},X_{5},X_{6}\}$ would place it transverse to the world volume of the brane while taking $\{X_{1},X_{5},X_{6}\}$ would place it within the world volume of the brane (i.e. by setting $X_{2}=0$ on the brane).  In both cases there is an obvious $SO(2)$ symmetry in the choice of the orientation of the giant magnon, or $U(1)$ symmetry of the magnon state.  These two cases are now discussed separately.
\nl\nl
The string that moves on $\{X_{3},Z\}$ was argued to be equivalent to a sine-Gordon soliton scattering from a boundary with a Neumann condition.  That the string has to meet the brane at $|Y|=X_{3}=0$ forces the end point to move on the equator of it's $S^{2}$, which is the edge of the disc at $X_{3}=0$ representing the brane.  The string's coordinates at the end point on the brane will obey
\bea
X_{3}'|_{x=0}\neq 0,\quad~Z'|_{x=0}=0\\
\dot{X_{3}}|_{x=0}=\ddot{X}_{3}|_{x=0}=0,\quad\dot{Z}|_{x=0}\neq 0.
\eea
Differentiating the relation for the sine-Gordon field $\varphi$ from \eqref{SGDef} we get
\bea
\varphi'{\rm sin}\frac{\varphi}{2}{\rm cos}\frac{\varphi}{2}\Big|_{x=0}&=&2X_{i}'X_{i}''|_{x=0},\quad\qquad i=3,5,6\nn\\
&=&-2X_{i}'\ddot{X}_{i}|_{x=0}\\
&=&0
\eea
where in the middle step we used the sigma model equations of motion
\be
\ddot{X}_{i}-X''_{i}+{\rm cos}(\varphi)X_{i}=0
\ee
along with the condition $X_{i}X'_{i}=0$ to avoid evaluating $X_{i}''|_{x=0}$.  The conclusion is that $\varphi'(x,t)|_{x=0}=0$; a Neumann condition.

To construct the corresponding string solution we must find the values of $\alpha_{+}$ and $\beta$ that result from the Neumann sine-Gordon condition.  The integrable boundary parameters take the values $M=0$ and $\varphi_{0}=2m\pi,~m\in\mathbb{Z}$ in this case, or in terms of $\zeta$ and $\eta$ (equations \eqref{zetaeta}), $\zeta=0$ and $\eta=\frac{\pi}{2}$.  As discussed in section 2, the Neumann condition requires that the discrete parameter $\epsilon=-1$, meaning that a soliton reflects as an anti-soliton.  The value of the sine-Gordon phase delay $a_{+}$ is then set by \eqref{a} to be
\be
a_{+}^{{\rm Neumann}}=2{\rm ln}\left\{{\rm cos}\frac{p}{2}\right\}\label{aNeumann}
\ee
which means a time delay for boundary scattering $\Delta T_{B}$ of
\be
\Delta T_{B}=\frac{{\rm sin}\frac{p}{2}}{{\rm cos}\frac{p}{2}}a_{+}=2\frac{{\rm sin}\frac{p}{2}}{{\rm cos}\frac{p}{2}}{\rm ln}\left\{{\rm cos}\frac{p}{2}\right\}.\label{DeltaTBY0}
\ee
Via the dictionary \eqref{aplus} we have
\be
\alpha_{+}^{{\rm Neumann}}={\rm ln}\left\{(1-{\rm sin}\frac{p}{2})^{2}~{\rm cos}^{2}\frac{p}{2}\right\}.
\ee
The value of $\beta$, determined through \eqref{betaphi0} depends upon the boundary conditions at $x\to\infty$ (if we are defining the half line to be $0\leq x$) but will always be divergent in the Neumann case because $\beta=\beta(\varphi_{0})$ controls the position of the central soliton, which for a Neumann boundary condition must be pushed off to infinity.  So for example, with $\varphi(x\to\infty)=0$ and $\epsilon_{1}=-1$, i.e. we send in an anti-kink, then
\be
\beta^{{\rm Neumann}}\to+\infty.
\ee

That we dispose of the central soliton this way means that we could treat this solution using only a two soliton solution of the string or sine-Gordon theories, and it was as such that this problem was considered in \cite{HofmanII}. The leading contribution to the magnon boundary scattering phase was calculated and found to be
\be
\delta_{B,~{\rm Neumann}}^{{\rm Max}~Y=0}=-8g{\rm cos}\frac{p}{2}{\rm ln}\left\{{\rm cos}\frac{p}{2}\right\}.\label{deltaNeumannHM}
\ee
The sine-Gordon phase delay on the other hand was given above in \eqref{aNeumann} which if used to calculate a $\delta_{B}$ through the prescription \eqref{deltaB} in fact produces an extra term on top of the Hofman and Maldacena result \eqref{deltaNeumannHM}.  Namely we get

\be
\delta_{B,~{{\rm Neumann}}}^{{\rm Max}~Y=0}=-8g{\rm cos}\frac{p}{2}{\rm ln}\left\{{\rm cos}\frac{p}{2}\right\}+\delta'\label{deltaY0}
\ee
where
\be
\delta'=8g{\rm cos}\frac{p}{2}.
\ee
The extra term is lacking in \cite{HofmanII} as the time delay $\Delta T_{B}$ the authors use is modified in order to present their results in the same language as those coming from the Bethe ansatz\cite{Arutyunov:2004vx}, making clear an ambiguity in the definition of the magnon S-matrix explained in \cite{Hofman:2006xt}.

\vspace{0.5cm}

The second maximal $Y=0$ case consists giant magnons contained within the world volume of the brane, on coordinates $\{X_{1},Z\}$.  As such the string end point is free, and by \eqref{SGDef} we have a Dirichlet sine-Gordon boundary condition with $\varphi_{0}=2m\pi,~m\in\mathbb{Z}$.  Similarly to the first maximal $Y=0$ case we find that $\beta\to\pm\infty$ (depending upon our exact choice of boundary conditions).  If we choose to send in a soliton from $x=\infty$ at $t\to-\infty$ then $\varphi_{0}=2\pi$ gives, by \eqref{aDirichlet},
\be
a_{+}^{{\rm Dirichlet},~\varphi_{0}=2\pi}=2{\rm ln}\left\{{\rm cos}\frac{p}{2}\right\},
\ee
so that both maximal $Y=0$ cases return the same time delay, \eqref{DeltaTBY0}, and hence leading scattering phase, \eqref{deltaY0}.
\nl

\begin{figure}
\centering
\includegraphics[width=0.4\textwidth]{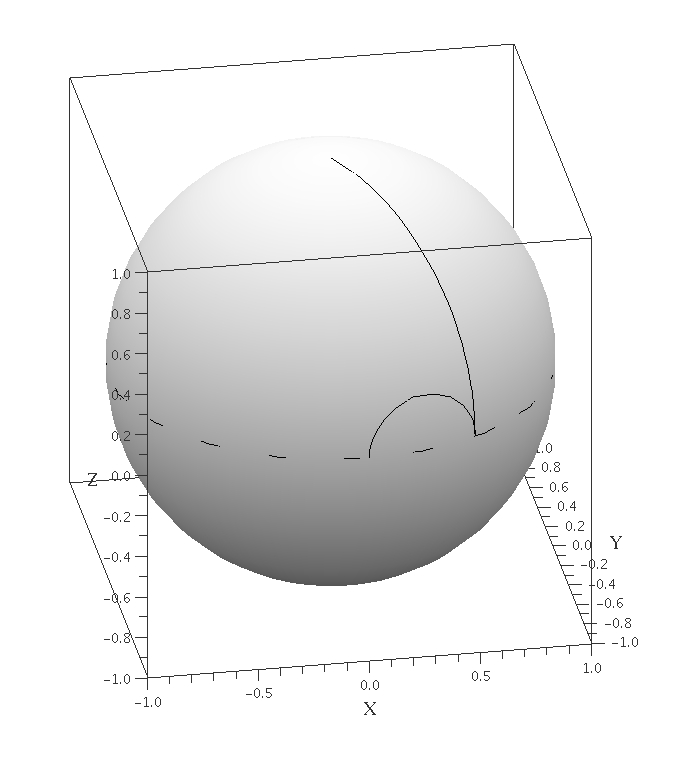}
\caption{A giant magnon attached to a maximal $Z=0$ giant graviton by the `boundary giant magnon' degree of freedom.\label{ZZeroEarly}}
\end{figure}

\noindent{\bf Maximal $Z=0$ case}
\nl\nl
This is the other orientation of a maximal giant graviton studied in \cite{HofmanII} where it was argued that the corresponding sine-Gordon picture is that of kinks or anti-kinks in reflection with a boundary with Dirichlet condition $\varphi_{0}=\pi$.  Considering the embedding of the maximal brane we have
\be
X_{1}^{2}+X_{2}^{2}+X_{3}^{2}+X_{4}^{2}=1\label{Z0}
\ee
so that by \eqref{S5} we have $|Z|=0$.  As always, the string moves on an $S^{2}\subset S^{5}$ of maximum radius 1, with two of its three directions being $Z=X_{5}+iX_{6}$.  We have a complete $SO(4)$ symmetry under the choice of the third coordinate so taking it to be $X_{3}$ the string coordinates must satisfy
\be
X_{3}^{2}+|Z|^{2}=1\label{Z0string}
\ee
and so in order that the string meet the $|Z|=0$ brane we have $X_{3}=\pm 1$.   This fixes the end point of the string to a single point at which $\dot{\vec{X}}^{2}|_{x=0}=0$, so that the definition \eqref{SGDef} tells us that the corresponding sine-Gordon boundary has the expected $\varphi_{0}=\pi$ Dirichlet condition.  Figure \ref{ZZeroEarly} depicts the string at early time.

Next we want to know what values of the moduli / boundary parameters to take in order to meet these boundary conditions.  We should expect $\alpha_{+}=\beta=0$ as this returns the solution without moduli that was considered implicitly in \cite{HofmanII}.  We know that at $t\to\pm\infty$ then $Z_{2}|_{x=0}=X_{3}|_{x=0}=\pm 1$ which from equations \eqref{CSminus} and \eqref{CSplus} immediately tells us that $\beta=0$.  To convince ourselves that the end point remains fixed to the pole for all time we can demand, for instance, from \eqref{N1A} that $Z_{1}|_{x=0}=0$ by taking $\tilde{N}_{1}|_{x=0}=0~\forall ~t$.  Putting $\beta=0$ in \eqref{N1A} gives
\bea
\tilde{N}_{1}|^{\beta=0}_{x=0}=-8sc\epsilon_{1}{\rm sinh}(v\gamma t){\rm sinh}\left(\frac{\alpha_{+}}{2}\right)\hspace{3cm}\nn\\
\hspace{2cm}+i\left[\epsilon_{0}(1-s)^{2}{\rm sinh}(\alpha_{+})-\epsilon_{0}(1+s)^{2}{\rm sinh}(\alpha_{+})\right].\label{N1zero}
\eea
Equation \eqref{aplus} relates $\alpha_{+}$ to the sine-Gordon phase delay as $\alpha_{+}=2{\rm ln}(1-s)-a_{+}$ while equation \eqref{aDirichlet} for $a_{+}^{{\rm Dirichlet}}$ with $\varphi_{0}=\pi$ then gives us
\be
\alpha_{+}^{{\rm SG-Dirichlet}}|_{\varphi_{0}=\pi}=0.
\ee
Equation \eqref{N1zero} then gives $\tilde{N}_{1}=0~\forall~t$ as required.  So indeed $\alpha_{+}=\beta=0$ for the maximal $Z=0$ case.

\begin{figure}
\centering
\begin{tabular}{cc}
\includegraphics[width=0.3\textwidth]{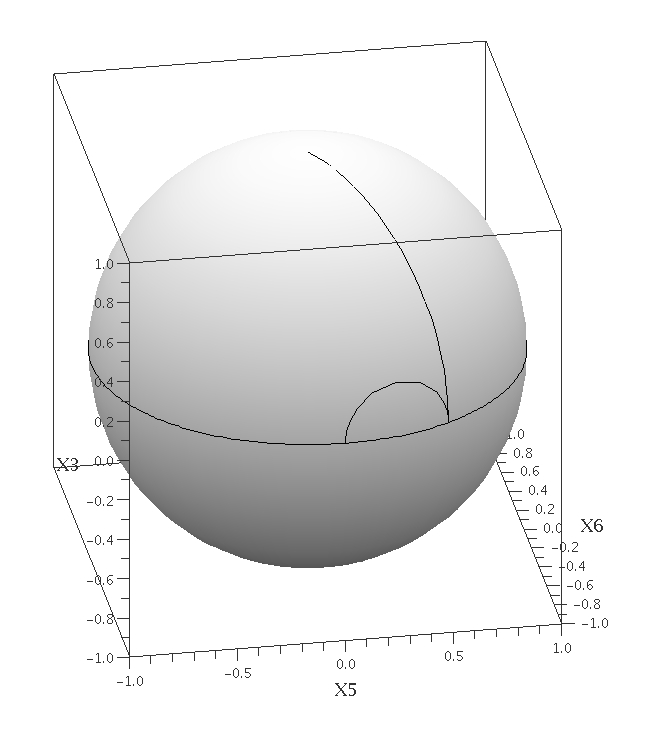} &
\includegraphics[width=0.3\textwidth]{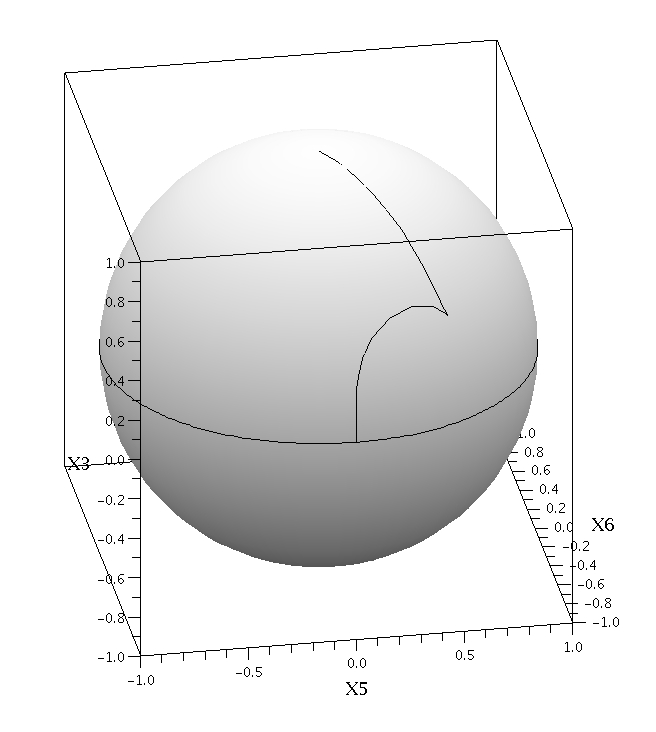} \\
\includegraphics[width=0.3\textwidth]{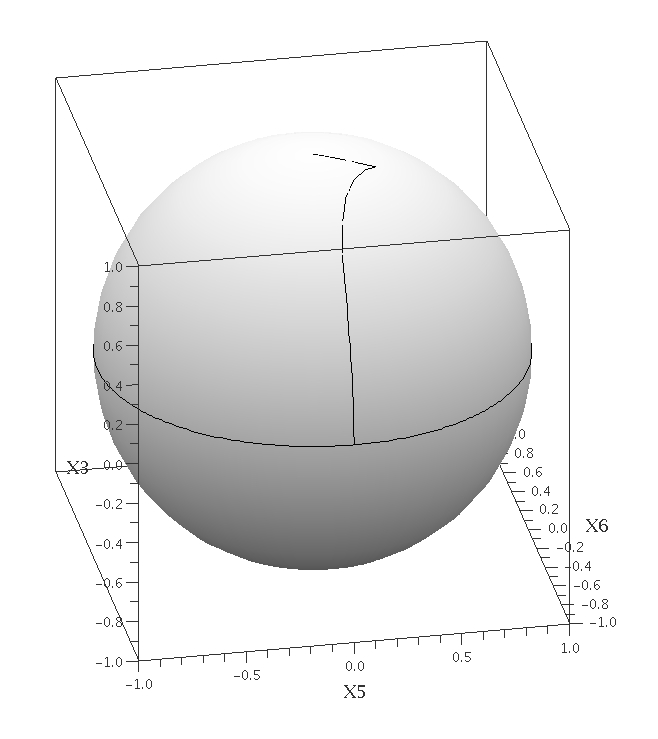} &
\includegraphics[width=0.3\textwidth]{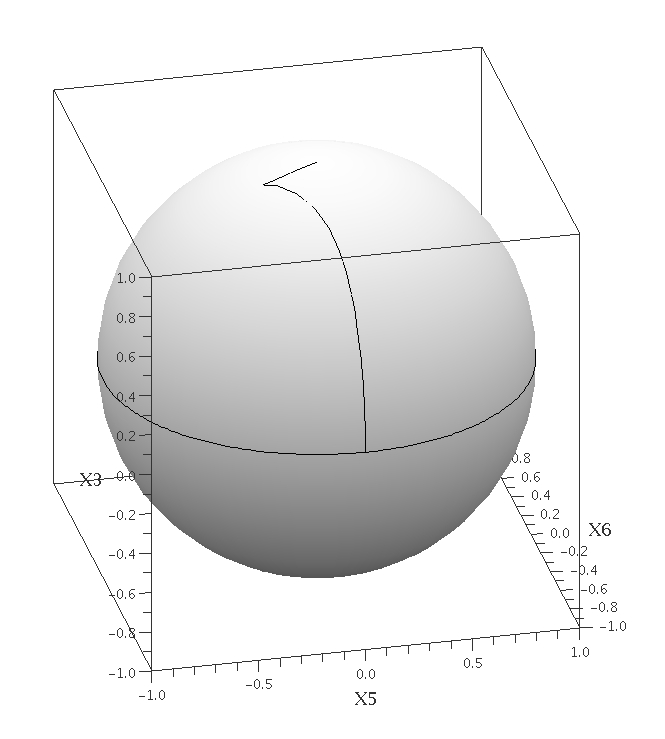} \\
\includegraphics[width=0.3\textwidth]{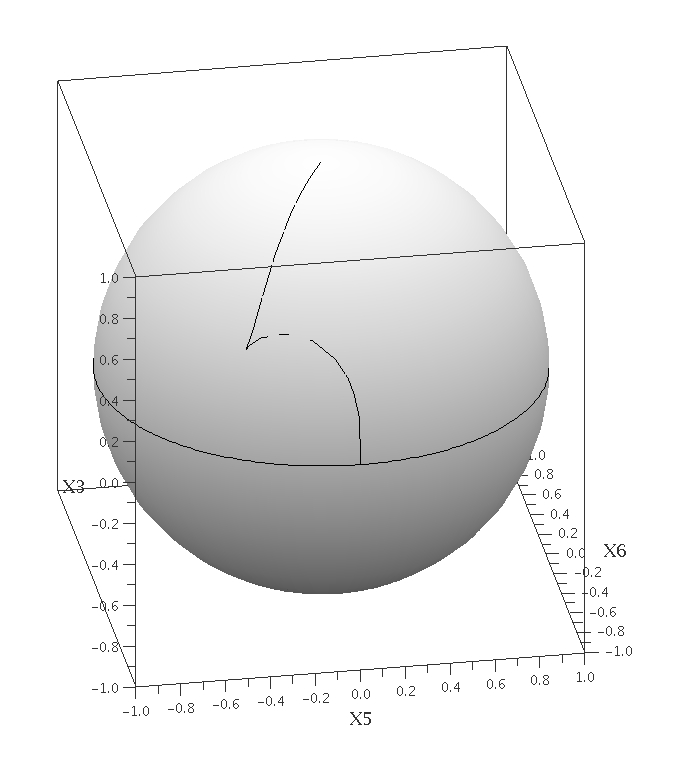} &
\includegraphics[width=0.3\textwidth]{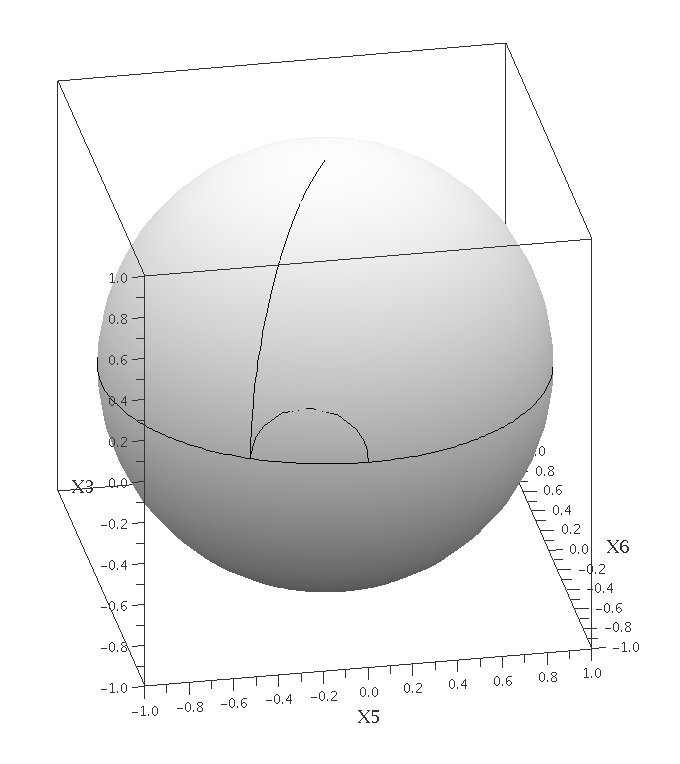}
\end{tabular}
\caption{This sequence shows the scattering of a giant magnon of momentum $p=\frac{\pi}{6}$ with a maximal $Z=0$ giant graviton in the co-moving frame, starting at $t=-20$ at the top left and proceeding to $t=+20$ at the lower right.  The cusp that sits on the equator at early times swings around the pole at which the brane is located and returns to the equator at late times, in a mirrored configuration.\label{ZZeroSequence}}
\end{figure}

Figure \ref{ZZeroSequence} depicts the scattering of a giant magnon off the maximal $Z=0$ giant graviton.  At early and late times we have a giant magnon attached at one end to a boundary giant magnon\cite{HofmanII,Bak:2008xq,Ciavarella:2010je}, an arc of the string that connects the equator to the brane that intersects the sphere at the pole.  At finite time an excitation spins the string about the sphere keeping its end point attached to the pole and returning the configuration to the mirror image (in the co-moving frame) of the $t\to-\infty$ configuration.

The leading contribution to the magnon boundary scattering phase as calculated in \cite{HofmanII} is
\be
\delta_{B}^{{\rm Max}~Z=0}=-4g{\rm cos}\frac{p}{2}{\rm ln}\left\{{\rm cos}\frac{p}{2}\left(\frac{1-{\rm sin}\frac{p}{2}}{1+{\rm sin}\frac{p}{2}}\right)\right\}.
\ee
The sine-Gordon phase delay on the other hand is given by \eqref{aDirichlet} with $\varphi_{0}=\pi$, giving
\be
a(p)^{{\rm Max}~Z=0}={\rm ln}\left\{(1-{\rm sin}\frac{p}{2})^{2}\right\}={\rm ln}\left\{{\rm cos}^{2}\frac{p}{2}\left(\frac{1-{\rm sin}\frac{p}{2}}{1+{\rm sin}\frac{p}{2}}\right)\right\}.
\ee
If we obtain a $\delta_{B}$ from this expression using \eqref{deltaB} we again get extra terms on top of the Hofman and Maldacena result,
\be
\delta_{B}^{{\rm Max}~Z=0}=\delta_{B}^{{\rm HM}}+\delta'
\ee
where now
\be
\delta'=8g{\rm cos}\frac{p}{2}+8g\left(\frac{p}{2}-\frac{\pi}{2}\right).
\ee
We have the same term $-8g{\rm cos}\frac{p}{2}$ as appeared in the $Y=0$ cases along with a constant phase shift of $-8g\frac{\pi}{2}$ and a term proportional to $p$, which is associated with the presence of the central soliton.  The term may be understood as follows.  Consideration in \cite{Hofman:2006xt} of the bulk scattering of two magnons with momenta $p_{1}$ and $p_{2}$ gave a scattering phase
\be
\delta(p_{1},p_{2})=-4g\left({\rm cos}\frac{p_{1}}{2}-{\rm cos}\frac{p_{2}}{2}\right){\rm ln}\left\{\frac{{\rm sin}^{2}\frac{p_{1}-p_{2}}{4}}{{\rm sin}^{2}\frac{p_{1}+p_{2}}{4}}\right\}-4gp_{1}{\rm sin}\frac{p_{2}}{2}
\ee
obtained by integration of the time delay for the scattering of two bulk sine-Gordon solitons along the lines of \eqref{deltaB}.  This contains a term proportional to $p_{1}$ that does not appear in the large $\lambda$ limit of the results of \cite{Arutyunov:2004vx} and is associated with a difference in the measure of length given to the soliton on the world sheet.  Our term above appears as the equivalent of this term with $p_{2}=\pi$, which is the momentum of the central soliton.

\subsection{Scattering with non-maximal $Y=0$ giant gravitons}

Among the pairs of integrable sine-Gordon boundary parameters $\{\zeta,\eta\}$ we are able to find a family that produce string solutions we associate with open strings ending upon non-maximal $Y=0$ giant gravitons.

We describe the non-maximal $Y=0$ case where the giant magnon is transverse to the world volume of the brane.  A $Y=0$ giant graviton of general radius $\rho$ is described by the relations
\be
|W|^{2}+|Z|^{2}=\rho^{2},\qquad 0\leq \rho\leq 1.\label{S3}
\ee
for the world volume coordinates and
\be
|Y|^{2}+\rho^{2}=1
\ee
for the embedding of the brane into $S^{5}$.  As $\rho$ is a constant and our string that lives on the 2-sphere is described by
\be
X_{3}^{2}+|Z|^{2}=1\label{S2}
\ee
then the string must meet the brane at a constant $X_{3}=|Y|=\pm\sqrt{1-\rho^{2}}$.

\begin{figure}
\centering
\includegraphics[width=0.4\textwidth]{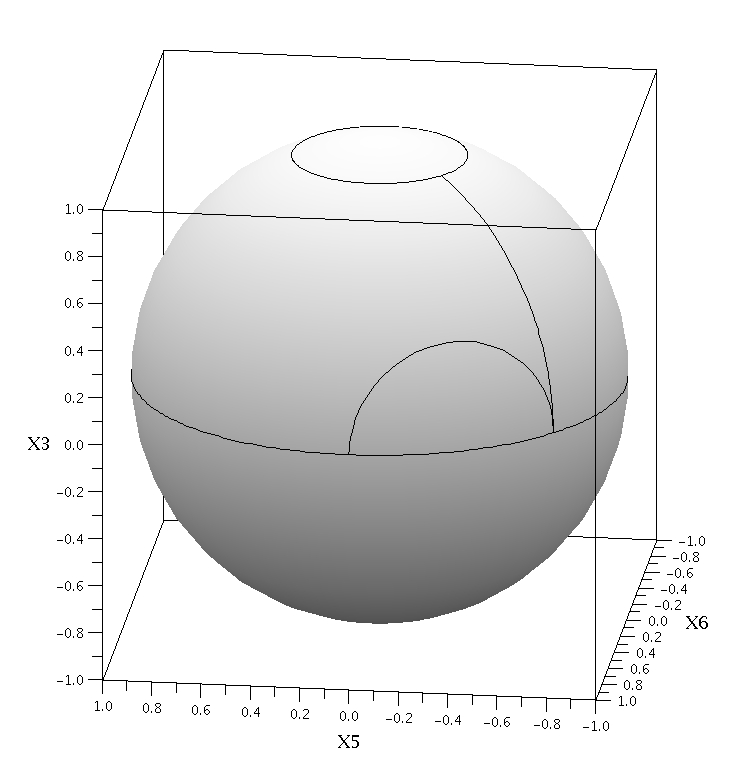}
\caption{A giant magnon attached to a non-maximal $Y=0$ giant graviton whose world volume is transverse to the string, at early time.  The giant magnon is attached via a boundary giant magnon analogously to the maximal $Z=0$ case.\label{YZeroEarly}}
\end{figure}

\begin{figure}
\centering
\begin{tabular}{ccc}
\includegraphics[width=0.3\textwidth]{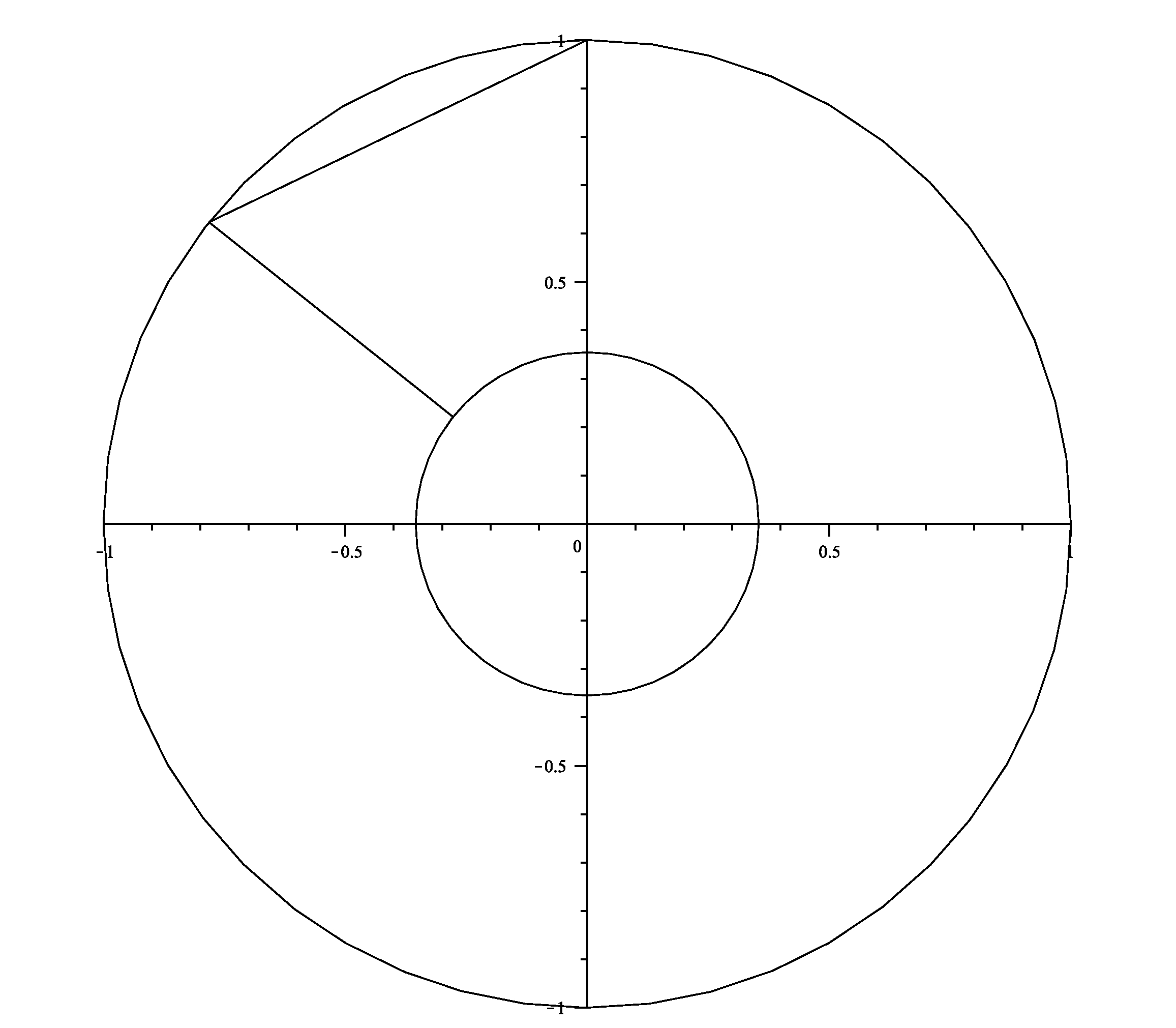} &
\includegraphics[width=0.3\textwidth]{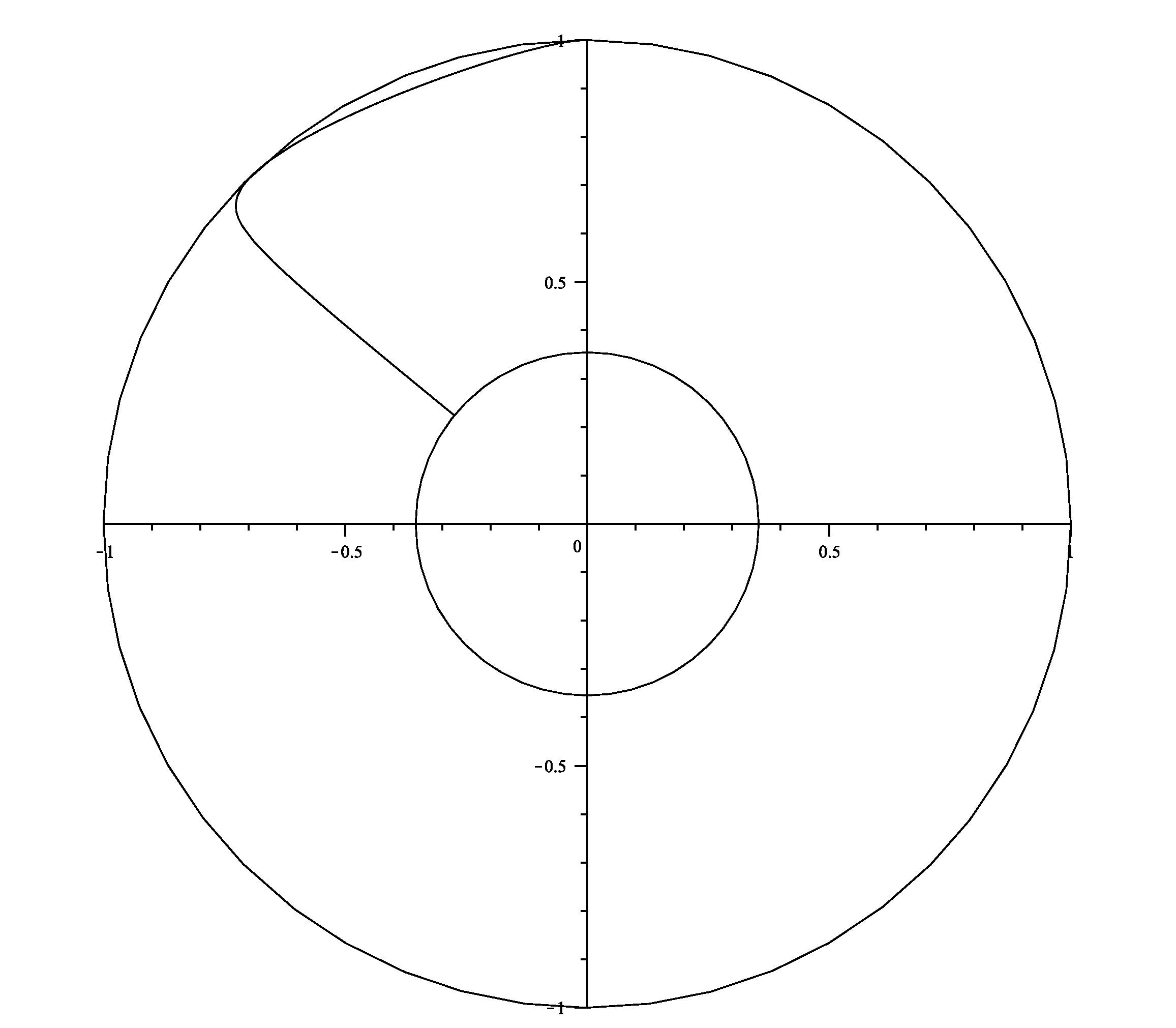} &
\includegraphics[width=0.3\textwidth]{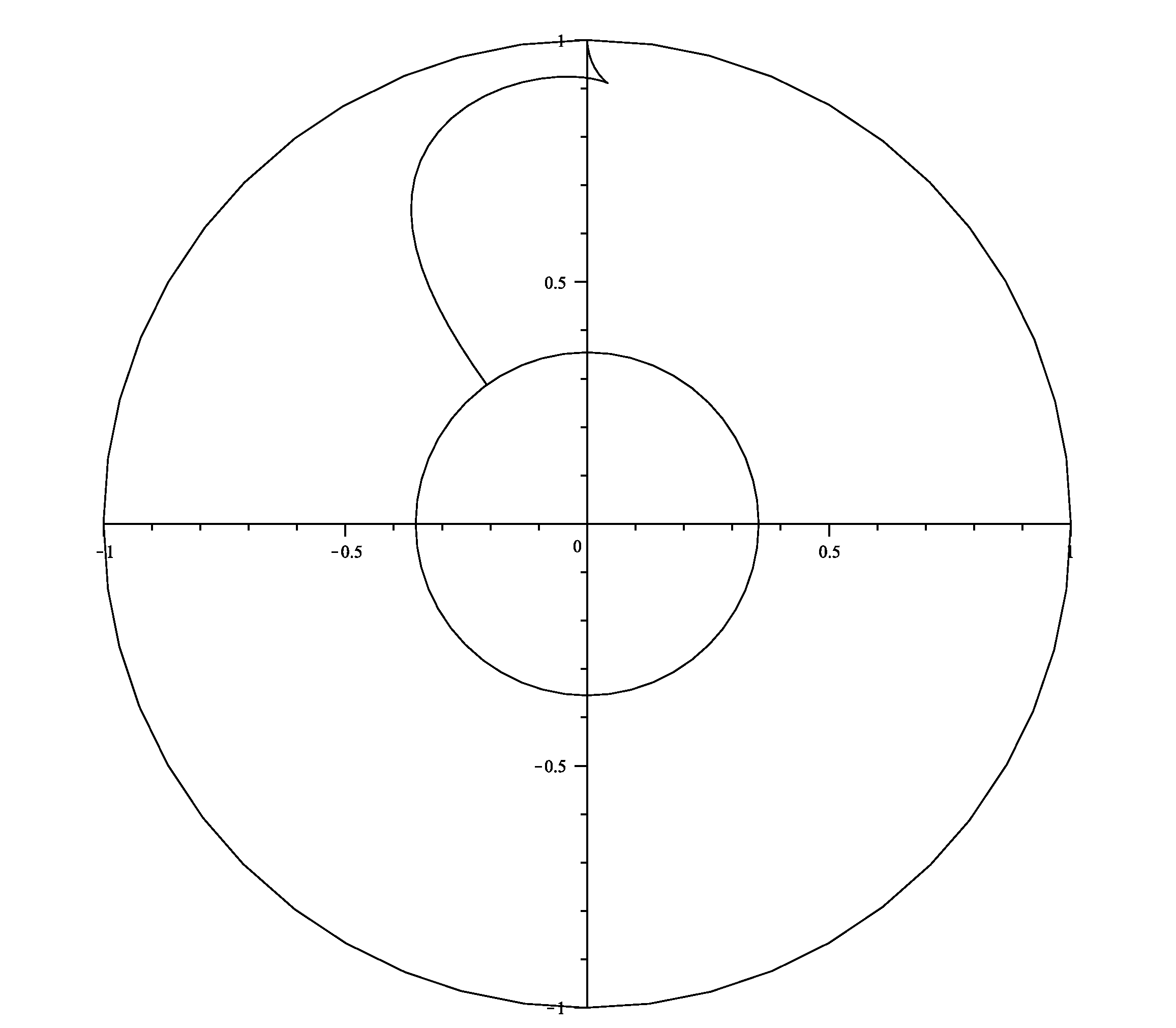} \\
\includegraphics[width=0.3\textwidth]{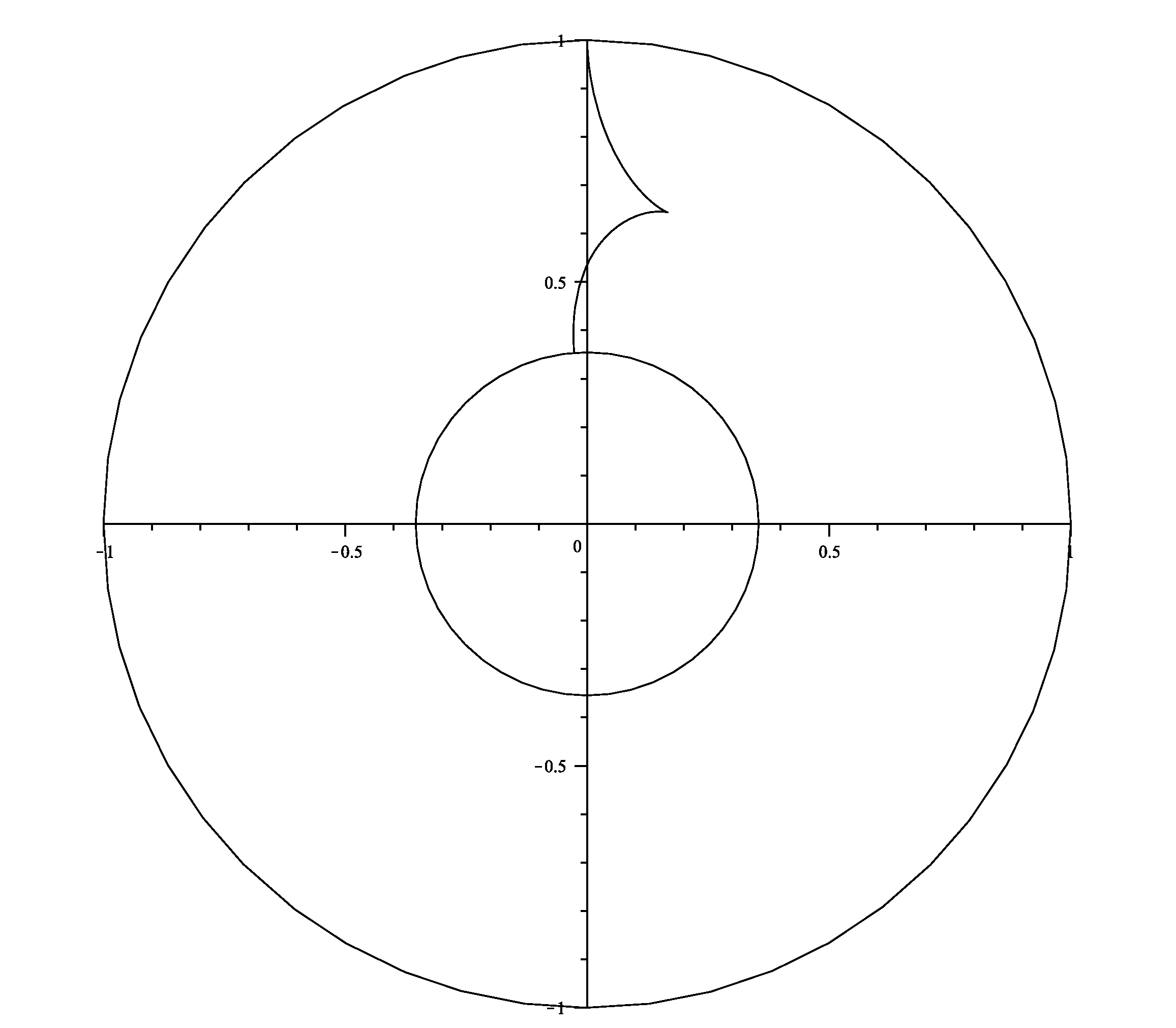} &
\includegraphics[width=0.3\textwidth]{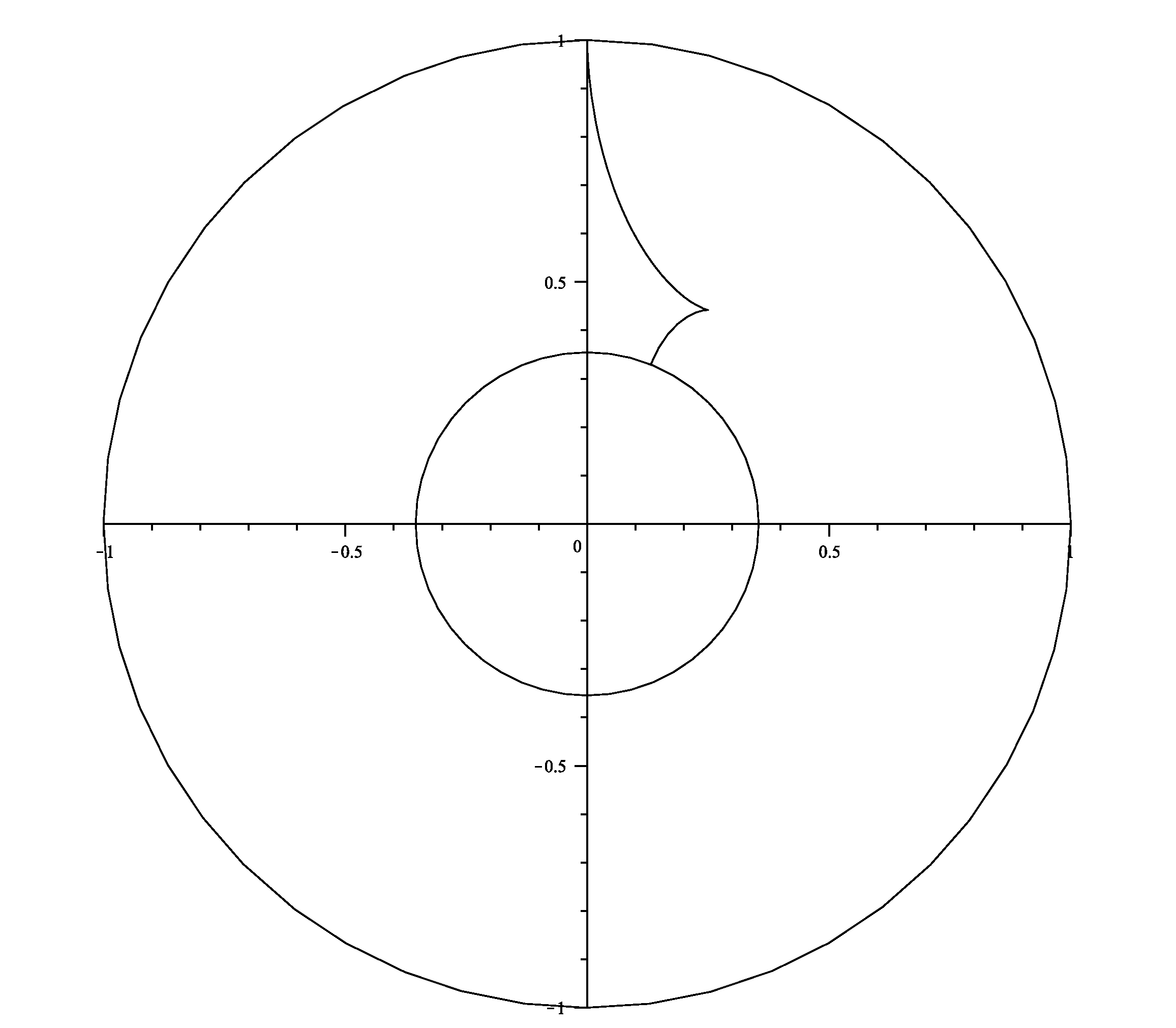} &
\includegraphics[width=0.3\textwidth]{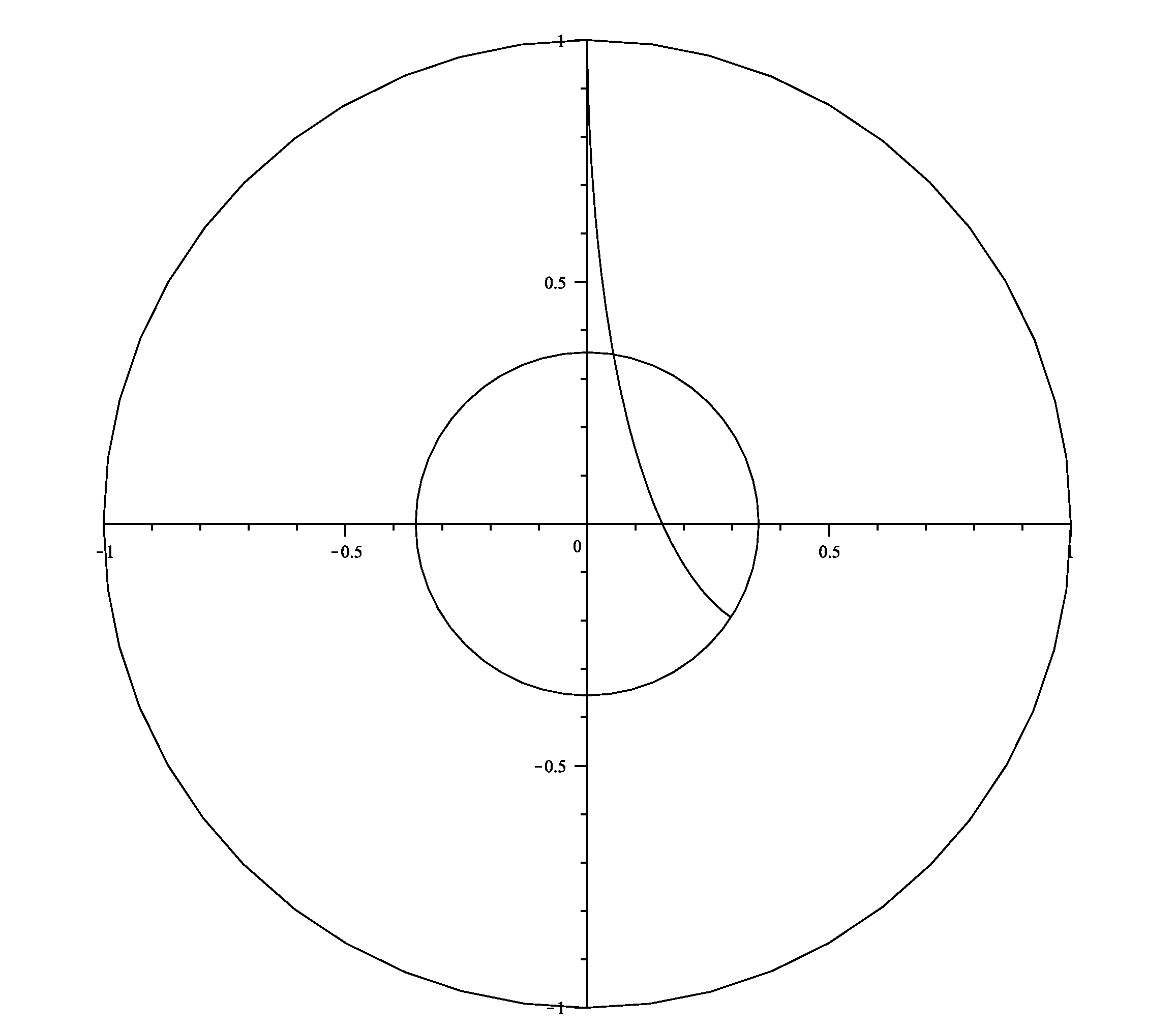} \\
\includegraphics[width=0.3\textwidth]{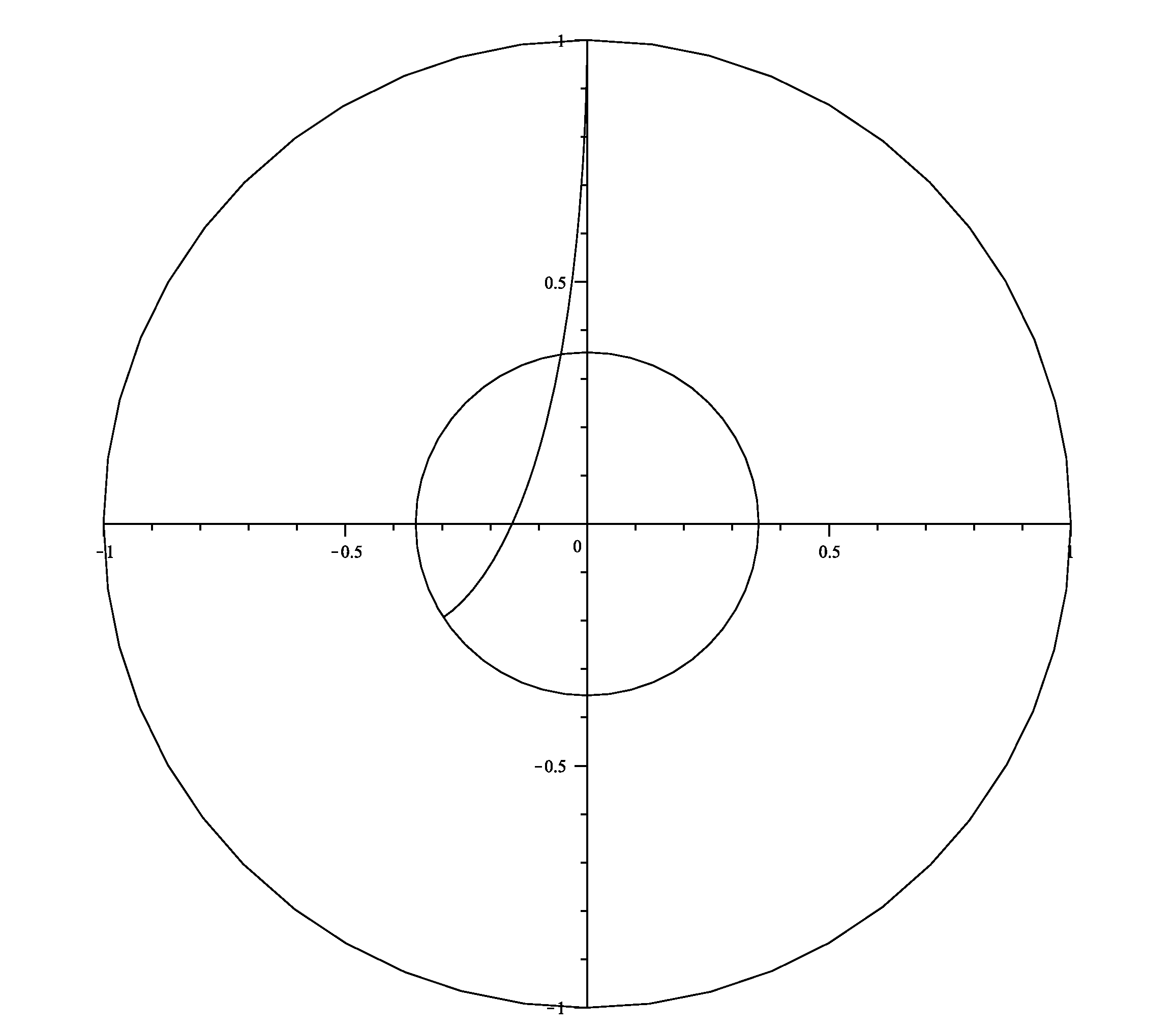} &
\includegraphics[width=0.3\textwidth]{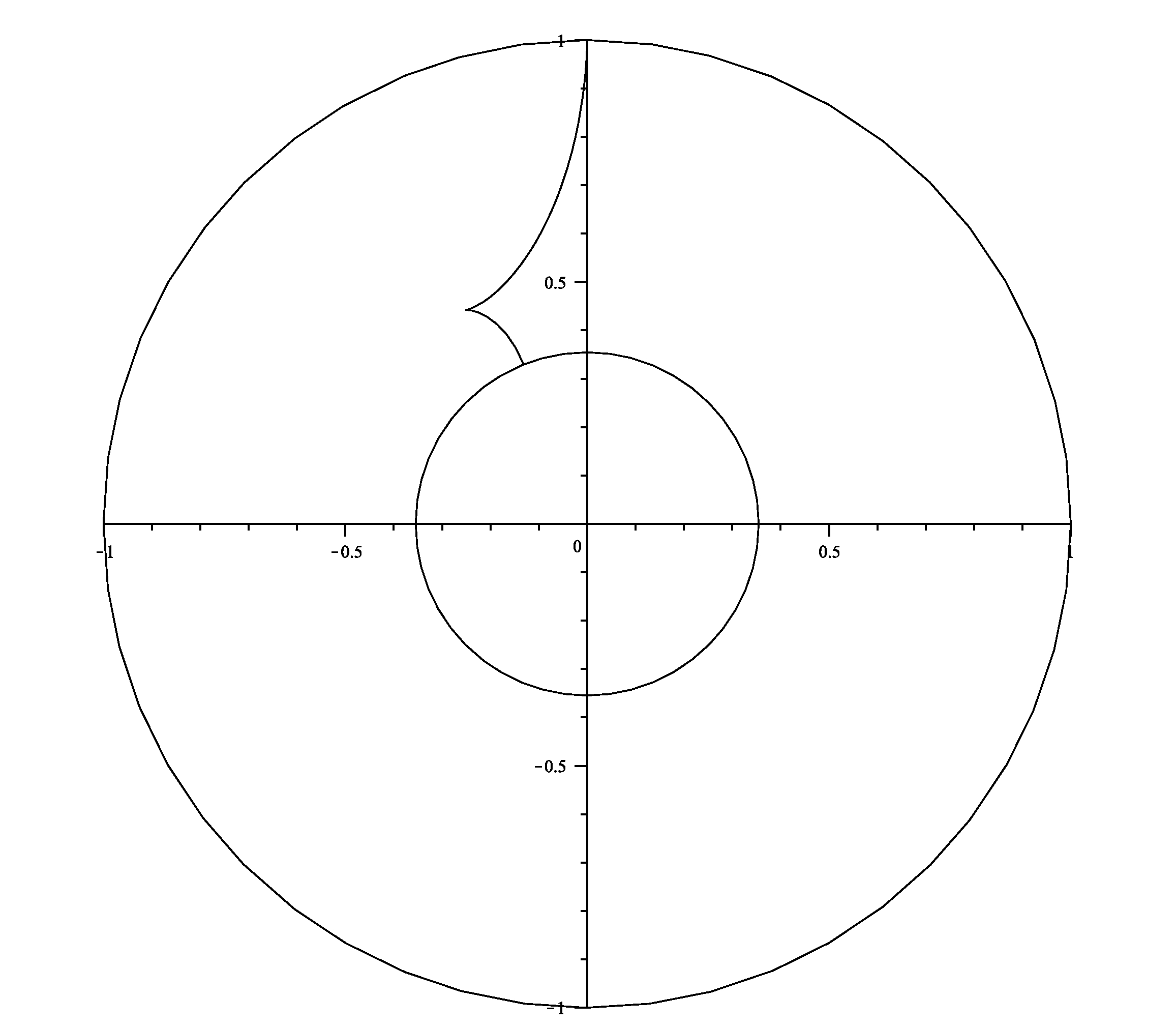} &
\includegraphics[width=0.3\textwidth]{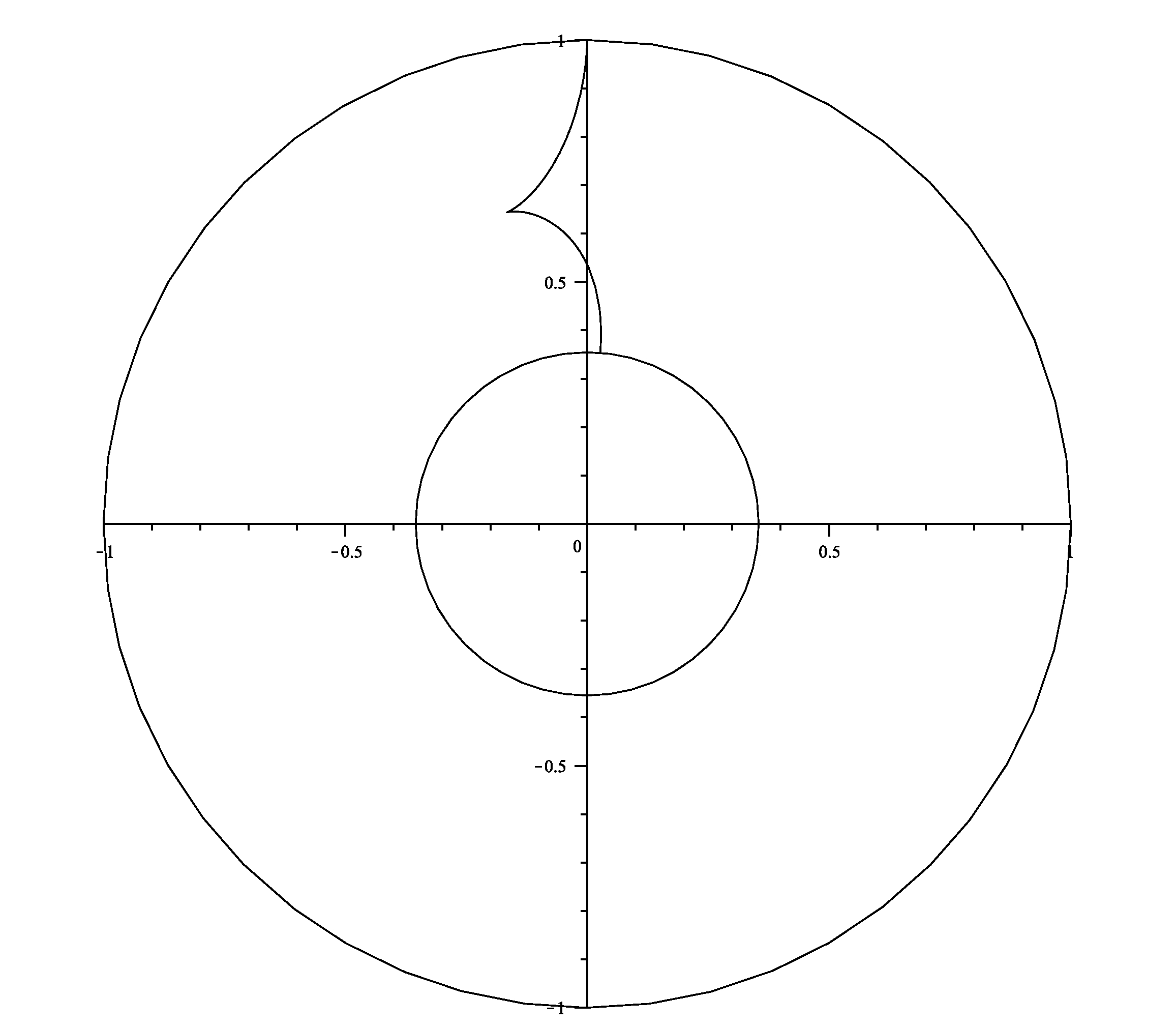} \\
\includegraphics[width=0.3\textwidth]{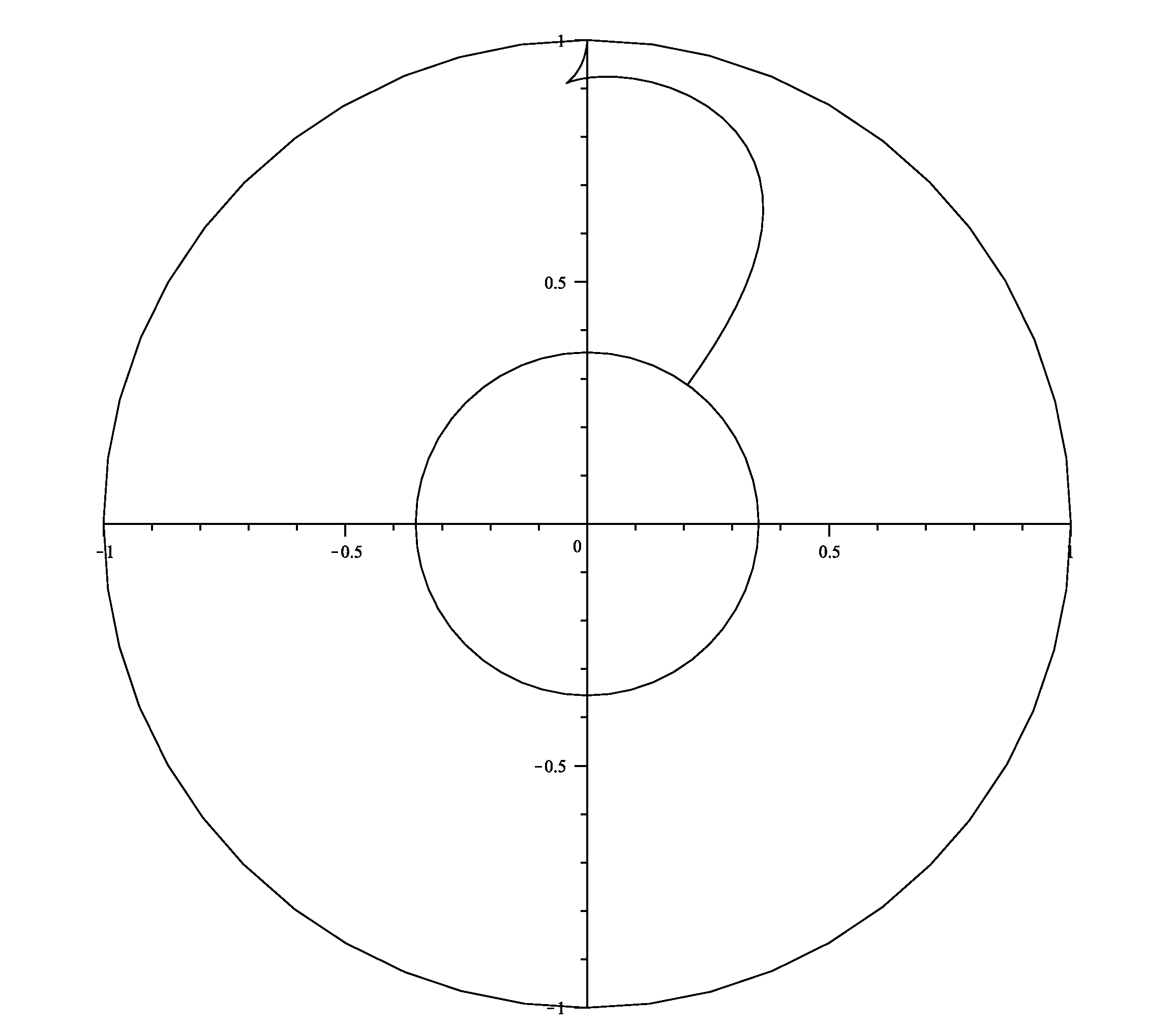} &
\includegraphics[width=0.3\textwidth]{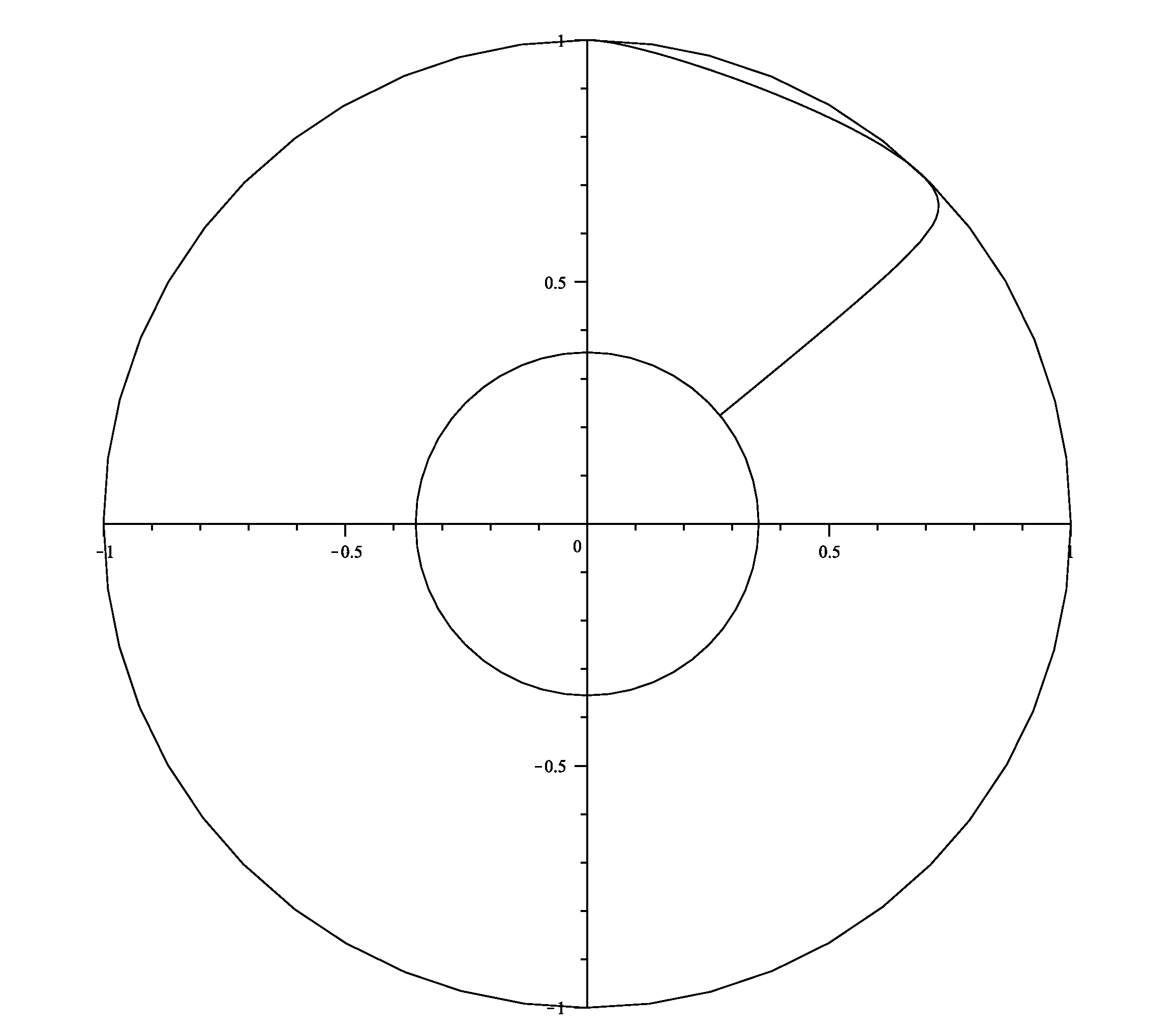} &
\includegraphics[width=0.3\textwidth]{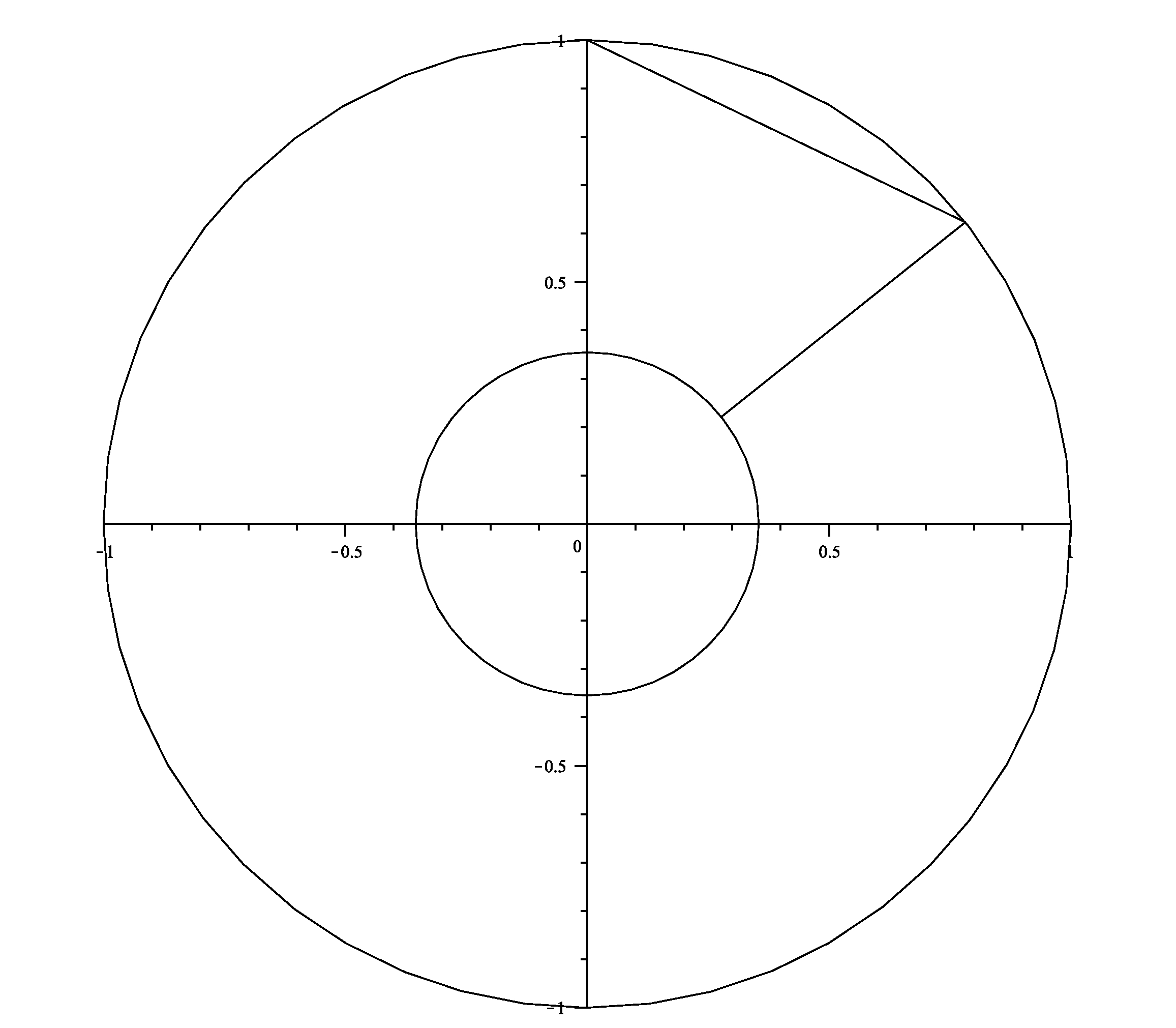}
\end{tabular}
\caption{This sequence shows the scattering of a giant magnon with a non-maximal $Y=0$ giant graviton projected to the $Z$ plane in which the graviton forms a disc of less than unit radius whose edge is depicted by a circle.  At the top left we start with $t=-20$ and at the bottom right we have $t=+20$.\label{YZeroSequence}}
\end{figure}

In the maximal case $\rho=1$ the brane appears, as before, as a unit disc located at $X_{3}=0$ and the string must meet it at the edge of the disc, $|Z|=1$.  For $0<\rho<1$ this disc shrinks in radius but the string must still end at the edge of the disc.  At early and late times then we will find the development of a boundary giant magnon\cite{Bak:2008xq,Ciavarella:2010je} connecting the scattering giant magnon ending at the equator $|Z|=1$ with the brane located at $|Z|=\rho$.  See Figure \ref{YZeroEarly}.  This non-maximal $Y=0$ case then begins to resemble the $Z=0$ case in which we encountered a boundary degree of freedom.  Figure \ref{YZeroSequence} depicts a scattering process with a giant magnon projected into the $Z$-plane.

We therefore seek a relation between the parameters of the string solution of the previous section such that $X_{3}|_{x=0}=Z_{2}|_{x=0}$ is a constant.  This value will be equal to that of $Z_{2}|_{x=0}$  when we take $t\to\pm\infty$,
\be
Z_{2}(x=0,t\to\pm\infty)=\frac{-\epsilon_{2}}{{\rm cosh}(\beta)}.\label{Z2const}
\ee
By equating the time dependent solution (with $x=0$) to this value and separately arranging for the coefficients of ${\rm cosh}(v\gamma t)$ and ${\rm cosh}(2v\gamma t)$ to vanish $\forall~t$ we obtain a trivial identity and the relations
\bea
{\rm tanh}\left(\frac{\alpha_{+}}{2}\right)&=&\frac{{\rm tanh}(\beta)}{s},\quad{\rm for}\quad\epsilon=+1\label{Circular1}\\
{\rm coth}\left(\frac{\alpha_{+}}{2}\right)&=&\frac{{\rm tanh}(\beta)}{s},\quad{\rm for}\quad\epsilon=-1.\label{Circular2}
\eea
The string end point will live on a circle in the $Z$-plane with radius $|Z|=\rho$ which from \eqref{Z2const} must satisfy
\be
\rho=\sqrt{1-\frac{1}{{\rm cosh}^{2}(\beta)}}=|{\rm tanh}(\beta)|
\ee
Then we will have a value of $\alpha_{+}$, related to the scattering time delay, for a given magnon momentum $p$ and giant graviton radius $\rho$.

Note that the value of $\Delta-J$ for this solution is a constant.  The end point at $x=0$ remains always in a radial orientation when projected into the $Z$-plane and so it is straightforward that we have zero flux of angular momentum $J$ off the string end point.  Indeed, if we had started by demanding
\be
\dot{J}=j|^{x=\infty}-j_{x=0}=0,
\ee
where $j$ is the angular momentum current density, along with satisfaction of the Virasoro constraint $\dot{X}_{i}X_{i}'=0$ then we would find ourselves asking for the same condition as produced \eqref{Circular1} and \eqref{Circular2}.  We have
\be
\Delta-J=\frac{\sqrt{\lambda}}{\pi}\left\{{\rm sin}\frac{p}{2}+\sqrt{1-\rho^{2}}\right\}
\ee
which is equal to the sum of the values for the scattering giant magnon and the boundary giant magnon separately at early or late time.
\nl

The relations \eqref{Circular1} and \eqref{Circular2} indicate that ${\rm sin}\frac{p_{{\rm c}}}{2}={\rm tanh}(\beta)$ is a critical momentum and by entering these values into the solution \eqref{ZA} - \eqref{N2A} we find that for a given size of non-maximal giant $\rho$ the giant magnons with momentum $p_{{\rm c}}$ are such that the string finds itself perfectly balanced across the pole of $S^{2}$ mid way through the scattering process.  As such the time delay for the scattering diverges.  From a sine-Gordon method of images perspective this divergence is due to the incoming soliton passing behind the boundary at $x=0$, as depicted in Figure \ref{TimeDelay}, and taking a large amount of time to scatter with the mirror soliton which we then find emerging from the boundary again as the string finally slips off the pole.  When the time delay diverges the boundary value of the sine-Gordon field, $\varphi(x,t)|_{x=0}$ must therefore fluctuate by a full $2\pi$ from it's asymptotic value of $\varphi_{0}$ during the process.

We would like to see this divergence from the expression for the sine-Gordon phase delay $a_{+}$ given in \eqref{a}.  We will also want to consider this quantity to calculate the semi-classical phase shift for the magnons so to these ends we first re-express \eqref{a}.

Using
\be
v^{2}=1-s^{2},\qquad \kappa^{2}=\frac{1-s}{1+s},\quad s\equiv {\rm sin}\frac{p}{2},
\ee
we can write
\be
a_{+}(p)={\rm ln}\left\{-\epsilon(1-s)^{2}\left[\left(\frac{1-s~{\rm cos}(\eta)}{1+s~{\rm cos}(\eta)}\right)\left(\frac{1+s~{\rm cosh}(\zeta)}{1-s~{\rm cosh}(\zeta)}\right)\right]^{\pm 1}\right\}.\label{azetaeta}
\ee
Our relations \eqref{Circular1} and \eqref{Circular2} for circular end-point motion may then be used to eliminate one of the boundary parameters, although we must translate from $\{\alpha_{+},\beta\}$ to $\{\zeta,\eta\}$ or $\{M,\varphi_{0}\}$, which in fact for our particular uses may be achieved very neatly as follows.

Selecting the positive power in \eqref{azetaeta} and using the relation between $a_{+}$ and $\alpha_{+}$ \eqref{aplus} we find
\bea
{\rm tanh}\left(\frac{\alpha_{+}}{2}\right)&=&-\frac{{1-s^{2}{\rm cosh}(\zeta){\rm cos}(\eta)}}{s({\rm cosh}(\zeta)-{\rm cos}(\eta))},\quad{\rm for}\quad \epsilon=+1\\
{\rm tanh}\left(\frac{\alpha_{+}}{2}\right)&=&-\frac{s({\rm cosh}(\zeta)-{\rm cos}(\eta))}{1-s^{2}{\rm cosh}(\zeta){\rm cos}(\eta)},\quad{\rm for}\quad\epsilon=-1
\eea
which are the inverse of one another.  Then relating $\beta$ to $\varphi_{0}$ by \eqref{betaphi0} we find
\be
{\rm tanh}(\beta)={\rm cos}\left(\frac{\varphi_{0}}{2}\right)
\ee
so that for both $\epsilon=\pm 1$ the conditions \eqref{Circular1} and \eqref{Circular2} for a circular end point motion gives us the single equation
\be
-\frac{1-s^{2}{\rm cosh}(\zeta){\rm cos}(\eta)}{{\rm cosh}(\zeta)-{\rm cos}(\eta)}={\rm cos}\left(\frac{\varphi_{0}}{2}\right).\label{coshcos}
\ee
This can now be used to write the phase delay in terms of the single boundary parameter $\varphi_{0}$ and the momentum $p$.

Equation \eqref{azetaeta} may be expanded as
\be
a_{+}(p)={\rm ln}\left\{-\epsilon(1-s)^{2}\frac{1+s({\rm cosh}(\zeta)-{\rm cos}(\eta))-s^{2}{\rm cosh}(\zeta){\rm cos}(\eta)}{1-s({\rm cosh}(\zeta)-{\rm cos}(\eta))-s^{2}{\rm cosh}(\zeta){\rm cos}(\eta)}\right\}.
\ee
The definition
\be
M{\rm cos}\left(\frac{\varphi_{0}}{2}\right)=2{\rm cosh}(\zeta){\rm cos}(\eta)\label{Mc}
\ee
then allows us to substitute immediately for the final terms of the numerator and denominator inside the log.  The lone ${\rm cosh}(\zeta)$ and ${\rm cos}(\eta)$ would normally result in an untidy expression but in this instance we may use \eqref{coshcos} and \eqref{Mc} to write
\be
{\rm cosh}(\zeta)-{\rm cos}(\eta)=-\frac{1-\frac{1}{2}s^{2}M{\rm cos}\left(\frac{\varphi_{0}}{2}\right)}{{\rm cos}\left(\frac{\varphi_{0}}{2}\right)}
\ee 
and, cancelling a common factor involving $M$, we find the sine-Gordon phase delay for the boundary condition corresponding to giant magnons reflecting from $Y=0$ giant gravitons can be written simply as 
\be
a_{+}(p)^{Y=0}={\rm ln}\left\{-\epsilon\left(1-{\rm sin}\frac{p}{2}\right)^{2}\left(\frac{{\rm cos}\left(\frac{\varphi_{0}}{2}\right)-{\rm sin}\frac{p}{2}}{{\rm cos}\left(\frac{\varphi_{0}}{2}\right)+{\rm sin}\frac{p}{2}}\right)\right\}.\label{aYzero}
\ee

This expression must interpolate to the case applicable to the scattering of giant magnons off maximal $Y=0$ giant gravitons with $\rho=1$, which corresponds to a Neumann boundary condition in sine-Gordon theory.  From the discussion above we have
\be
\rho=r=|{\rm tanh}(\beta)|=\Big|{\rm cos}\left(\frac{\varphi_{0}}{2}\right)\Big|
\ee
and so the phase delay in terms of the graviton radius and magnon momentum is equivalently
\be
a_{+}(p)^{Y=0}={\rm ln}\left\{-\epsilon\left(1-{\rm sin}\frac{p}{2}\right)^{2}\left(\frac{\rho+{\rm sin}\frac{p}{2}}{\rho-{\rm sin}\frac{p}{2}}\right)\right\}.\label{aYzero}
\ee
When ${\rm sin}\frac{p}{2}<\rho$ we should take $\epsilon=-1$, a Neumann-like solution, and when ${\rm sin}\frac{p}{2}>\rho$ take $\epsilon=+1$, a Dirichlet-like solution.  When $\rho=1$ we get
\be
a_{+}(p)^{{\rm Maximal}~Y=0}=2{\rm ln}\left\{{\rm cos}\frac{p}{2}\right\}
\ee
which is indeed the Neumann phase delay produced above.

We can now calculate the semi-classical phase shift for the scattering of magnons with the non-maximal giant graviton by the integral
\be
\delta_{B}^{Y=0}=4g\int\frac{s}{\sqrt{1-s^{2}}}{\rm ln}\left\{-\epsilon(1-s)^{2}\left(\frac{\rho+s}{\rho-s}\right)\right\}{\rm d}s.
\ee
Putting this into the form of the maximal, or Neumann, contribution and a term appearing when $\rho<1$ we could write
\be
\delta_{B}^{Y=0}=4g\int\left[\frac{2s}{\sqrt{1-s^{2}}}{\rm ln}\left\{\sqrt{1-s^{2}}\right\}+\frac{s}{\sqrt{1-s^{2}}}{\rm ln}\left\{\left(\frac{1-s}{1+s}\right)\left(\frac{\rho+s}{\rho-s}\right)\right\}\right]{\rm d}s
\ee
and the answer that results is
\be
\delta_{B}^{Y=0}=-4g{\rm cos}\frac{p}{2}{\rm ln}\left\{{\rm cos}^{2}\frac{p}{2}\left(\frac{1-{\rm sin}\frac{p}{2}}{1+{\rm sin}\frac{p}{2}}\right)\left(\frac{\rho+{\rm sin}\frac{p}{2}}{\rho-{\rm sin}\frac{p}{2}}\right)\right\}+\delta'.\label{deltanonmax}
\ee
with
\be
\delta'=8g{\rm cos}\frac{p}{2}+8g\left(\frac{p}{2}-\frac{\pi}{2}\right).
\ee

The extra factors that appear within the log for $\rho<1$ coincide with the appearance of the boundary degree of freedom we have described.  In fact it is the same factor we saw in the maximal $Z=0$ case but dressed with an extra factor containing the parameter $\rho=\Big|{\rm cos}\left(\frac{\varphi_{0}}{2}\right)\Big|$.  If we were to trust the $\rho\to 0$ limit\footnote{In the $\rho\to 0$ limit the curvature of the giant graviton of course becomes large so that corrections to the supergravity description would generally become necessary.} when the $Y=0$ brane shrinks to zero size then the boundary giant magnon present at early and late times extends all the way to the pole of $S^{2}$ giving a string solution identical to the maximal $Z=0$, or $\varphi_{0}=\pi$ sine-Gordon Dirichlet, case with an identical scattering phase to match.  For $\rho=1$ of course we recover the maximal $Y=0$ case.

The extra terms $\delta'$ appear from the integrals associated with the pure Neumann piece and the undressed boundary piece, and are not to do with the added integrable boundary condition.  Indeed, the extra terms are associated with the direct scattering of the method of images solitons while the new factor appearing in \eqref{deltanonmax}, or \eqref{aYzero}, represents simply the insertion of an added time delay to the scattering process such that the scattering of the in-going soliton and the mirror out-going soliton occurs some way behind the point $x=0$, consistent with the origin of the modulus $\alpha_{+}$ in the initial displacement parameters of the bulk solution.  Seen as an added time delay $\tilde{\Delta T}$ we can see immediately from \eqref{aYzero} that this quantity is given by
\be
\tilde{\Delta T}=\frac{{\rm sin}\frac{p}{2}}{{\rm cos}\frac{p}{2}}{\rm ln}\left\{\frac{\rho+{\rm sin}\frac{p}{2}}{\rho-{\rm sin}\frac{p}{2}}\right\}.
\ee

\subsection{Other open strings with integrable sine-Gordon boundary conditions}

It is interesting to ask what string solutions correspond to other particular integrable sine-Gordon boundary conditions.  For general $\{\zeta,\eta\}$ the string end point will execute some motion that we do not associate with any obvious configuration of a D-brane embedded in $S^{5}$.  We will also generically have $\dot{J}\neq 0$ so that the brane must accept some angular momentum from the string, albeit temporarily.
\nl

\noindent{\bf Strings with sine-Gordon Dirichlet boundary conditions}
\nl

We might ask about the family of sine-Gordon Dirichlet boundary conditions, of which we have seen two, these representing all those strings whose end points have a constant speed (by \eqref{SGDef}).  It is possible to show that all such strings possess end points that, while maintaining a constant speed, execute circular motions in the co-moving frame with angular velocity in the $Z$-plane $\dot{\theta}=1$.  The angle of the axis of this circle with respect to the $Z$-plane is $\frac{\varphi_{0}}{2}$.

We prove this claim as follows.  If we select the co-moving frame and rotate our solution by an angle of $\frac{\pi}{2}-\frac{\varphi_{0}}{2}$ toward the vertical so that the circle upon which the end point moves is horizontal then we may proceed as in section 4.2 to obtain a relation between the moduli $\alpha_{+}$ and $\beta$ by demanding that the value of the new $Z_{2}(x=0)$ be constant for all time.  We obtain
\be
{\rm tanh}\left(\frac{\alpha_{+}}{2}\right)={\rm sin}\frac{p}{2}{\rm cos}\frac{\varphi_{0}}{2}
\ee
which is exactly the form of ${\rm tanh}\left(\frac{\alpha_{+}}{2}\right)$ we find by taking the form of $\alpha_{+}$ due to the sine-Gordon Dirichlet phase delay \eqref{aDirichlet} (with positive power) and dictionary \eqref{aplus}.

We note that as ${\rm tanh}(\beta)={\rm cos}\frac{\varphi_{0}}{2}$ this relation is very similar to the condition \eqref{Circular1} for motion on the non-maximal $Y=0$ brane when $\epsilon=+1$, as it must necessarily be in the Dirichlet case.

It would be tempting to think of this co-moving circle as a brane configuration intersecting $S^{2}$ analogously to the non-maximal $Y=0$ giant of section 4.2 but rotated at an angle to the axis of large $J$.  As such however the string end point would not meet the brane orthogonally (as is seen immediately at early and late times).
\nl

\noindent{\bf Scattering with non-maximal $Z=0$ giant gravitons?}
\nl

For the case of a non-maximal $Z=0$ giant graviton the string end point must be some point away from the pole.  For the radius of the graviton $\rho\neq 1$ then a BPS brane must sit on a circle in the $Z$-plane, some $0< |Z|\leq 1$ with $|Z|=\sqrt{1-\rho^{2}}$.  This point now moves at a constant speed of $\dot{\theta}\sqrt{1-\rho^{2}}$ where we must know $\dot{\theta}$, the angular velocity about the circle in the $Z$-plane, or about the $J$-axis.  We can show that $\dot{\theta}=1$ (when the radius of $S^{5}$ is $R=1$) as follows.

In \cite{McGreevy:2000cw} McGreevy et al. used the Born-Infeld (plus Chern-Simons potential) effective action for giant gravitons wrapping $S^{3}\subset S^{5}$ with radius $0\leq \rho \leq 1$ to show that they possess angular momentum $J$ given by
\be
J=N\rho^{3}\frac{\dot{\theta}(1-\rho^{2})}{\sqrt{1-\dot{\theta}^{2}(1-\rho^{2})}}+N\rho^{4}
\ee
where $N$ is here the number of units of 5-form flux on $S^{5}$ (equivalently the number of colours in the gauge theory dual).  Note $J\propto N$. Rearranging this expression for $\dot{\theta}$ we have
\be
\dot{\theta}=\frac{\left(\frac{J}{N}-\rho^{4}\right)}{\sqrt{\rho^{6}(1-\rho^{2})^{2}+\left(\frac{J}{N}-\rho^{4}\right)^{2}(1-\rho^{2})}}.
\ee
If we then use the BPS condition for the radius of the brane at fixed $J$,
\be
\rho^{2}=\frac{J}{N},
\ee
then we find
\be
\dot{\theta}=1.
\ee
Now, supposing that in the large $N$ limit the string has only angular momentum of order $\sqrt{N}$ to give to the brane during scattering then we expect the brane to remain `heavy', and to be unperturbed by the scattering of the giant magnon.  The string end point must remain fixed to this point which proceeds about the circle $|Z|=\sqrt{1-\rho^{2}}$ at $\dot{\theta}=1$.

In the co-moving frame with $\dot{\theta}=1$ then the string end point must remain completely fixed.  Using the solution \eqref{Z1} - \eqref{N2general} it is then easy to see that we cannot describe such an end point motion using the method of images.

\section{Discussion}

We have described the scattering of giant magnons with certain maximal and non-maximal giant gravitons by applying the method of images to obtain solutions to the boundary string sigma model from the bulk 3-soliton solution.  We have also calculated the corresponding overall scattering phases at large $\lambda$ through knowledge of the time delays for scattering in the sine-Gordon picture.  As such we have addressed a question about the status of string solutions resulting from the less trivial integrable boundary conditions of the Pohlmeyer reduced string asked in \cite{HofmanII}, and contributed to the understanding of the map between sine-Gordon theory and the sigma model.

In the most interesting case a giant magnon of arbitrary momentum $p$ is scattered from a non-maximal $Y=0$ giant graviton whose world volume is entirely transverse to the string.  It is seen that as we reduce the radius $\rho$ of the giant graviton from its maximal value $\rho=1$ we find the development of a boundary degree of freedom, or boundary giant magnon, akin to that found in the maximal $Z=0$ case.  In fact the large $\lambda$ boundary scattering phase is seen to resemble that obtained for scattering with a maximal $Z=0$ giant but for a factor that interpolates appropriately between the  maximal $Y=0$ case without a boundary giant magnon and the maximal $Z=0$ case for which the boundary magnon stretches from the equator to the pole.  (However, this does not correspond to the rotation of a maximal $Y=0$ giant into a maximal $Z=0$ giant; the string configurations thus produced are simply identical in both cases.)

We were not able to describe the scattering of giant magnons with non-maximal $Z=0$ giant gravitons using this method and it seems likely that this process is not integrable.

It is interesting that we can find exact solutions for the scattering of giant magnons with non-maximal giant gravitons at all given that attempts to construct integrable boundary conditions along the lines of the bulk theory have previously failed\cite{Mann:2006rh}, and in fact it was found in \cite{MacKay:2004rz} the the $O(3)$ sigma model will only permit an integrable boundary along the lines of \cite{Mann:2006rh} where the string end point is constrained to move on a \emph{great} circle on $S^{2}$, consistent with some maximal giant gravitons but certainly not with a general non-maximal giant.  Nevertheless, we have shown that a special combination of the sine-Gordon integrable boundary parameters allows for an end point on a $Y=0$ non-maximal giant for which the Dirichlet sub-manifold of $S^{2}$ is an $S^{1}$ of radius $r=\rho<1$ whose axis is aligned with the angular momentum.

Clearly it is possible that among the various combinations of integrable boundary conditions in the sine-Gordon theory there may be described other brane configurations of interest and this is a direction for future thought.  It is possible that allowing the string to move on more of $AdS_{5}\times S^{5}$ will allow more freedom to meet certain non-maximal giant gravitons.  In particular an open string moving on $\mathbb{R}\times S^{3}$, which is already naturally described by the solutions used in this paper, corresponds under Pohlmeyer reduction to the Complex sine-Gordon theory on the half line\cite{Bowcock:2002vz} which has been explored in the literature.  We have also focussed purely on the string side of the correspondence and left consideration of the dual gauge theory operators for another time.

\vspace{1cm}
\noindent\large{\bf{Acknowledgements}}\newline\newline
\small
A. Ciavarella would like to thank P. Bowcock for guidance throughout this project and C. A. Young for discussion relevant to this work.

\appendix
\section{N=3 solution on $\mathbb{R}\times S^{2}$}

The notation here is that introduced at the start of sub-section 3.2 together with $\tilde{\zeta}_{k}\equiv \zeta_{k}-\alpha_{k}$.  The components $D$, $N_{1}$ and $N_{2}$ are to be composed as in \eqref{Z1} and \eqref{Z2}.

\bea
D=16e^{3it}\left\{-4s_{2}s_{3}(c_{1}-c_{2})(c_{1}-c_{3})\epsilon_{1}{\rm cosh}(\tilde{\zeta_{1}})\hspace{4cm}\right.\nn\\
\left.+4s_{1}s_{3}(c_{1}-c_{2})(c_{2}-c_{3})\epsilon_{2}{\rm cosh}(\tilde{\zeta_{2}})-4s_{1}s_{2}(c_{1}-c_{3})(c_{2}-c_{3})\epsilon_{3}{\rm cosh}(\tilde{\zeta_{3}})\right.\nn\\
\left.+\epsilon_{1}\epsilon_{2}\epsilon_{3}\left[(1-c_{1-2})(1-c_{1-3})(1-c_{2-3}){\rm cosh}(\tilde{\zeta_{1}}+\tilde{\zeta_{2}}+\tilde{\zeta_{3}})\right.\right.\hspace{2cm}\nn\\
\left.\left.+(1-c_{1-2})(1-c_{1+3})(1-c_{2+3}){\rm cosh}(\tilde{\zeta_{1}}+\tilde{\zeta_{2}}-\tilde{\zeta_{3}})\right.\right.\hspace{2cm}\nn\\
\left.\left.+(1-c_{1+2})(1-c_{1-3})(1-c_{2+3}){\rm cosh}(\tilde{\zeta_{1}}-\tilde{\zeta_{2}}+\tilde{\zeta_{3}})\right.\right.\hspace{2cm}\nn\\
\left.\left.+(1-c_{1+2})(1-c_{1+3})(1-c_{2-3}){\rm cosh}(\tilde{-\zeta_{1}}+\tilde{\zeta_{2}}+\tilde{\zeta_{3}})\right]\right\}\hspace{1.5cm}
\eea

\bea
N_{1}=16e^{3it}\left\{-4c_{1}s_{2}s_{3}(c_{1}-c_{2})(c_{1}-c_{3})\epsilon_{1}{\rm cosh}(\tilde{\zeta_{1}})\hspace{4cm}\right.\nn\\
\left.+c_{2}s_{1}s_{3}(c_{1}-c_{2})(c_{2}-c_{3})\epsilon_{2}{\rm cosh}(\tilde{\zeta_{2}})-4c_{3}s_{1}s_{2}(c_{1}-c_{3})(c_{2}-c_{3})\epsilon_{3}{\rm cosh}(\tilde{\zeta_{3}})\right.\nn\\
\left.+\epsilon_{1}\epsilon_{2}\epsilon_{3}\left[c_{1+2+3}(1-c_{1-2})(1-c_{1-3})(1-c_{2-3}){\rm cosh}(\tilde{\zeta_{1}}+\tilde{\zeta_{2}}+\tilde{\zeta_{3}})\right.\right.\nn\hspace{1cm}\\
\left.\left.+c_{1+2-3}(1-c_{1-2})(1-c_{1+3})(1-c_{2+3}){\rm cosh}(\tilde{\zeta_{1}}+\tilde{\zeta_{2}}-\tilde{\zeta_{3}})\right.\right.\nn\hspace{1cm}\\
\left.\left.+c_{1-2+3}(1-c_{1+2})(1-c_{1-3})(1-c_{2+3}){\rm cosh}(\tilde{\zeta_{1}}-\tilde{\zeta_{2}}+\tilde{\zeta_{3}})\right.\right.\nn\hspace{1cm}\\
\left.\left.+c_{-1+2+3}(1-c_{1+2})(1-c_{1+3})(1-c_{2-3}){\rm cosh}(\tilde{-\zeta_{1}}+\tilde{\zeta_{2}}+\tilde{\zeta_{3}})\right]\right.\hspace{0.5cm}\nn\\\nn\\
\left.+i\left[-4s_{1}s_{2}s_{3}\left((c_{1}-c_{2})(c_{1}-c_{3})\epsilon_{1}{\rm sinh}(\tilde{\zeta_{1}})\right.\right.\right.\nn\hspace{5cm}\\
\left.\left.\left.-(c_{1}-c_{2})(c_{2}-c_{3})\epsilon_{2}{\rm sinh}(\tilde{\zeta_{2}})+(c_{1}-c_{3})(c_{2}-c_{3})\epsilon_{3}{\rm sinh}(\tilde{\zeta_{3}})\right)\right.\right.\nn\\
\left.\left.+\epsilon_{1}\epsilon_{2}\epsilon_{3}\left[s_{1+2+3}(1-c_{1-2})(1-c_{1-3})(1-c_{2-3}){\rm sinh}(\tilde{\zeta_{1}}+\tilde{\zeta_{2}}+\tilde{\zeta_{3}})\right.\right.\right.\hspace{1cm}\nn\\
\left.\left.\left.+s_{1+2-3}(1-c_{1-2})(1-c_{1+3})(1-c_{2+3}){\rm sinh}(\tilde{\zeta_{1}}+\tilde{\zeta_{2}}-\tilde{\zeta_{3}})\right.\right.\right.\hspace{1cm}\nn\\
\left.\left.\left.+s_{1-2+3}(1-c_{1+2})(1-c_{1-3})(1-c_{2+3}){\rm sinh}(\tilde{\zeta_{1}}-\tilde{\zeta_{2}}+\tilde{\zeta_{3}})\right.\right.\right.\hspace{1cm}\nn\\
\left.\left.\left.+s_{-1+2+3}(1-c_{1+2})(1-c_{1+3})(1-c_{2-3}){\rm sinh}(-\tilde{\zeta_{1}}+\tilde{\zeta_{2}}+\tilde{\zeta_{3}})\right]\right]\right\}
\eea

\bea
N_{2}=32ie^{4it}\left.\Big\{-s_{1}s_{2}s_{3}\left[(1-c_{1+2})(1-c_{1-2})\right.\right.\nn\hspace{6cm}\\
\left.\left.+(1-c_{1+3})(1-c_{1-3})+(1-c_{2+3})(1-c_{2-3})\right]\right.\nn\hspace{1.5cm}\\
\left.+s_{1}(c_{1}-c_{2})(c_{1}-c_{3})\epsilon_{2}\epsilon_{3}\left[(1-c_{2-3}){\rm cosh}(\tilde{\zeta_{2}}+\tilde{\zeta_{3}})+(1-c_{2+3}){\rm cosh}(\tilde{\zeta_{2}}-\tilde{\zeta_{3}})\right]\right.\hspace{0.5cm}\nn\\
\left.-s_{2}(c_{1}-c_{2})(c_{2}-c_{3})\epsilon_{1}\epsilon_{3}\left[(1-c_{1-3}){\rm cosh}(\tilde{\zeta_{1}}+\tilde{\zeta_{3}})+(1-c_{1+3}){\rm cosh}(\tilde{\zeta_{1}}-\tilde{\zeta_{3}})\right]\right.\hspace{0.5cm}\nn\\
\left.+s_{3}(c_{1}-c_{3})(c_{2}-c_{3})\epsilon_{1}\epsilon_{2}\left[(1-c_{1-2}){\rm cosh}(\tilde{\zeta_{1}}+\tilde{\zeta_{2}})+(1-c_{1+2}){\rm cosh}(\tilde{\zeta_{1}}-\tilde{\zeta_{2}})\right]\right\}\hspace{0.4cm}
\eea

\end{document}